\documentclass[aps,pra,twocolumn,longbibliography,superscriptaddress,floatfix]{revtex4}

\usepackage{amsfonts}
\usepackage{amsthm}
\usepackage{amsmath}
\usepackage{amssymb}
\usepackage{mathtools}
\usepackage[all]{xy}
\input xy
\xyoption{matrix}
\xyoption{2cell}
\xyoption{arrow}
\xyoption{curve}
\xyoption{all}

\usepackage{graphicx}
\usepackage{epsfig}
\usepackage{dcolumn}% Align table columns on decimal point
\usepackage{bm}% bold math
\usepackage{wasysym}

\usepackage{amsmath}
\usepackage{graphicx}
\usepackage[latin1]{inputenc}
\usepackage{ulem}
\usepackage{epstopdf}
\usepackage{subfigure}
\usepackage{color}

\usepackage{graphicx}
\usepackage{epsfig}
\usepackage{dcolumn}% Align table columns on decimal point
\usepackage{bm}% bold math
\usepackage{amsmath}
\usepackage{graphicx}
\usepackage[latin1]{inputenc}
\usepackage{ulem}
\usepackage{epstopdf}
\usepackage{subfigure}
\usepackage{color}
\usepackage{amsthm}
\usepackage{newlfont}
\usepackage{graphicx}
\usepackage{epstopdf}
\usepackage{appendix}
\usepackage[breaklinks=true]{hyperref}
\usepackage{breakcites}
\usepackage{textcomp}
\usepackage{appendix}
\usepackage{multirow}	
\usepackage{color}
\usepackage{amssymb}
\usepackage{epsfig}
\usepackage{bm}
\usepackage[american]{babel}
\usepackage{braket}
\usepackage{soul}
\usepackage{float}
\hypersetup{colorlinks=true,linkcolor=blue,citecolor=blue,filecolor=blue,urlcolor=blue,pdfstartview=FitH}
\usepackage{textgreek}

\newcommand{\pcsadd}{Center for Theoretical Physics of Complex Systems, Institute for Basic Science, Daejeon 34126, Republic of Korea}

\begin{document}
\title{
Casting dissipative compact states in coherent perfect absorbers
%Coherent perfect absorption in multi-band lattice networks
}

\author{C. Danieli}
\email{cdanieli@pks.mpg.de}
\affiliation{Max Planck Institute for the Physics of Complex Systems,  Dresden D-01187, Germany}
\affiliation{\pcsadd}
\author{T. Mithun}
     \email{mthudiyangal@umass.edu}
\affiliation{Department of Mathematics and Statistics, University of Massachusetts, Amherst MA 01003-4515, USA}
\affiliation{\pcsadd}
%\affiliation{Center for Theoretical Physics of Complex Systems, Institute for Basic Science (IBS), Daejeon 34126, Republic of Korea}

\date{\today}

%%%%%%%%%%%%%%%%%%%%%%%%%%%%%%%%%%%%%%%%%%%
%%%%%%%%%%%%%%%             ABSTRACT                 %%%%%%%%%%%%%%
%%%%%%%%%%%%%%%%%%%%%%%%%%%%%%%%%%%%%%%%%%%
\begin{abstract}

Coherent perfect absorption (CPA), also known as time-reversed laser, is a wave phenomenon resulting from the reciprocity of destructive interference of transmitted and reflected waves. In this work we consider quasi one-dimensional lattice networks which posses at least one flat band, and show that CPA and lasing can be induced in both linear and nonlinear regimes of this lattice by fine-tuning non-Hermitian defects (dissipative terms localized within one unit-cell).  
We show that local dissipations that yield CPA simultaneously yield novel dissipative compact solutions of the lattice, whose growth or decay in time can be fine-tuned via the dissipation parameter. 
%This induces the existence of localized excitations that by simultaneously annihilating incoming and propagating radiation from the complex potential, emulate the later/absorber set up. 
The scheme used to numerically visualize the theoretical findings offers a novel platform for the experimental implementation of these phenomena in optical devices. 

\end{abstract}

\maketitle

%\section*{Introduction}

\begin{figure}[!htbp]
\centering
\includegraphics [width=0.8\columnwidth]{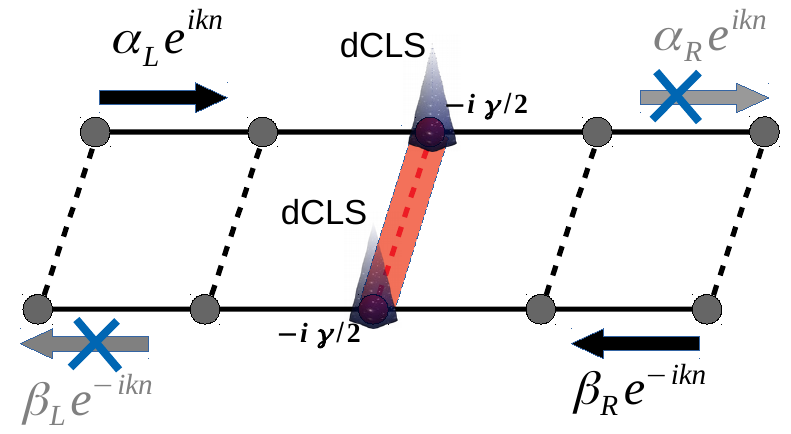}
\caption{
Schematic representation of CPA: $\alpha_L,  \beta_R$ and $\alpha_R, \beta_L$ control the incident and the reflected waves respectively; the red area represents the  localized dissipation.  
The trigones represent the dissipative compact state (dCLS).
}
\label{fig:CPA}
\end{figure}

The growing interest in non-Hermitian system is motivated by 
the novel and  unprecedented phenomena that gain and loss terms can generate.
The phenomenon of coherent perfect absorption (CPA) a notable discovery in this field which arises from the interplay of propagating waves in homogeneous media and local dissipations, and it is related to the concept of spectral singularity \cite{Mostafazadeh:2009,baranov2017coherent}. 
Introduced by Chong {\it et.al.} in Ref.\cite{Chong2010coherent} and also referred as time-reversed laser, two counter-propagating coherent radiations towards a local dissipation leads to the distructive-interference of the transmitted and the scattered waves upon the fine-tuning of the wave frequencies and the complex potential (see Fig.\ref{fig:CPA} for a schematic view of the process).
It was also noticed that by inverting the local dissipation into a gain potential - equivalently, reversing the time  - a perfect absorber can be turned into a laser where only outgoing radiations originated from the complex potential are present, while incoming ones are extinct \cite{Chong2010coherent}. 
%Notably, this perfect absorption phenomenon was experimentally realized in ...  \cite{}, and it has applications in ....  \cite{}.
Longhi \cite{Longhi:2010} and  Chong {\it et.al.} \cite{Chong:2011PT} realized that CPA and lasing can be simultaneously achieved, yielding laser-absorbers.  %in presence of $\mathcal{PT}$-symmetric potential, yielding laser-absorbers.
%These set-ups %, where both incoming and outgoing waves are at the same time absent, 
%were later extended to non $\mathcal{PT}$-symmetric systems \cite{Mostafazadeh_2012}. 
More recently, 
CPA and lasing phenomena have been extended to one-dimensional nonlinear lattice chains \cite{Zezyulin1018array,Mullerseaat6539} - where these phenomena are achieved in small propagating waves regimes - and nonlocal absorber \cite{jeffers2019nonlocal}.  %although they may suffer instabilities.} 
In particular, CPA was also realized experimentally in cold atomic systems \cite{Mullerseaat6539}. 
An important open frontier is to extend CPA to lattices which support multiple Bloch bands with distinct spectral properties.
This may induce additional novel phenomena in lattice in presence of a non-Hermitian source. 
Moreover, it would substantially extend the applicability and the experimental implementation of perfect absorbers and lasers.

In the recent years a growing amount of attention has been received by translationally invariant networks which 
simultaneously support propagating waves and compact localization. 
These systems  - commonly referred as flat band networks - are characterized by the existence of at least one dispersionless (flat) spectral band \cite{Leykam2018artificial,Leykam2018perspective}. 
The corresponding eigenmodes of the flat band are typically strictly compact in space - hence dubbed compact localized states (CLS) - and they arise due to destructive interference.  
Notable theoretical advances for these systems includes systematic generator schemes 
\cite{Flach2014detangling,Maimaiti2017compact,Rontgen2018compact,Maimaiti2019universal}
localization phenomena due to onsite perturbations 
\cite{Leykam2013flatband,Bodyfelt2014flatbands,Danieli2015flatbands,Leykam2017localization}, 
%non-hermitian homogeneous potentials \cite{Leykam2017flatbandNH},
and the existence of compact breathers in flat band networks with Kerr nonlinearity 
\cite{Johansson2015compactification,Ramachandran2018fano,Danieli2018compact}.  
Moreover, these systems have been 
realized experimentally in several set-ups, including 
photonics lattices  \cite{Mukherjee2015observation,Vicencio2015observation2,Weimann2016transport} 
to exciton-polariton condensates \cite{Masumoto2012exciton}
and ultra cold atoms \cite{Taie2015coherent}, among others.
These findings highlight how flat band networks are ideal set-ups to study and implement novel and highly relevant phenomena in condensed matter physics.

Both the CLS and CPA are wave phenomena resulting from destructive interference. However, the question {\it "can these phenomena be embedded in a single device?"} has not yet been tackled. 
In this work we address this yet unanswered question by considering flat band networks in presence of Kerr nonlinearity and complex non-Hermitian dissipative terms localized within one unit-cell. 
%In this work we address this question by consideing} samples of flat band networks with  {\color{red}  Kerr nonlinearity in presence of complex non-Hermitian dissipative terms localized within one unit-cell.}
We then analytically prove and numerically verify that the chosen dissipations can induce CPA phenomenon out of dispersive waves, as well as 
induce non-propagating excitations localized at the dissipative sites out of the flat band states. 

%\section*{Set-up: flat band networks, dissipation, and nonlinearity}

Let us consider the equations of motion of a network with $\nu$ sites per unit-cell 
\begin{equation}
i \dot{\Psi}_n = - H_0\Psi_n -  H_1\Psi_{n+1} - H_1^\dagger\Psi_{n-1} - \frac{i\gamma}{2}  V \Psi_0 \delta_{n,0}\
\label{eq:fb_eq}
\end{equation}
represented by the time-dependent  complex vector
 $\Psi_n \in\mathbb{C}^\nu$. 
Here $H_0$ and  $H_1$ are real square matrices of size $\nu$. 
The dissipation is localized within the unit-cell at $n=0$, where 
$\gamma > 0$ and the square matrix $V$ encodes the gain and loss terms of the non-Hermitian terms -  non-Hermitian defects. 
The steady-state solution $\Psi_n = U_n e^{-i \mu t}$ of Eq.(\ref{eq:fb_eq}) yields the eigenvalue problem
\begin{equation}
\mu U_n =  - H_0 U_n -  H_1 U_{n+1} - H_1^\dagger U_{n-1}-\frac{i\gamma}{2} V U_0 \delta_{n,0}.
\label{eq:fb_eq2}
\end{equation}
For $\gamma = 0$ (no dissipation),  
the Bloch solution $U_n = e^{ikn}\varPhi_k$ turns Eq.(\ref{eq:fb_eq2}) to $\mu \varPhi_k = B(k)  \varPhi_k $ where the matrix 
$B(k) \equiv - H_0  -  e^{ik}  H_1- e^{-ik}  H_1^\dagger$ depends on the wave vector $k$. 
This resulting eigenvalues of the matrix $B(k)$ yield the band structure formed by $\nu$ Bloch bands $\Omega = \{\mu_j(k)\}_{j=1}^\nu$ of Eq.(\ref{eq:fb_eq}). 
In this work we consider systems which contains at least one flat band. 
%
%Among these spectral bands, we assume that one band ({\it e.g.} the last one $\mu_\nu$) is independent on momentum $k$.
%For $\gamma \neq 0$, in order to induce CPA in Eq.(\ref{eq:fb_eq}) we 

Let us consider the following ansatz of solution for Eq.(\ref{eq:fb_eq})
\begin{equation}
   \begin{split}
    U_n & = \varPhi_j\left\{
                \begin{array}{ll}
                  \alpha_L e^{ikn}+\beta_L e^{-ikn} ~~~n < 0 \\
                  u_0 ~~~~~~~~~~~~~~~~~~~~~~~n=0  \\
                 \alpha_R e^{ikn}+\beta_R e^{-ikn} ~~~n > 0.
                \end{array}
              \right. 
  \label{eq:CS_ansatz}
     \end{split}
  \end{equation}
where $\varPhi_j$ is the eigenvector of $B(k)$ correspondent to the dispersive band $\mu_j$.
Hence, $\alpha_L$ and $\beta_R$ control the incident waves from the left ($L$), while $\alpha_R$ and $\beta_L$ the reflected wave from the right ($R$). 
In this framework, CPA implies to fine-tune the wave-vector $k$ in Eq.(\ref{eq:CS_ansatz}) with respect to the dissipation components 
$\{\gamma, V\}$ in order to admit the incident radiations $\alpha_L,  \beta_R$ and annihilate the reflected radiations $\alpha_R, \beta_L$ - as represented in Fig.\ref{fig:CPA}. 
In general, different choices of the dissipation matrix $V$ may be considered to induce CPA, depending on the dispersive wave considered. 
Certain choices of $V$ can both (i) induce fine-tuned CPA from counter-propagating dispersive waves 
and (ii) preserve the CLS of the flat band which overlaps with the dissipative cell, inducing non-propagating localized excitations centered at the dissipative unit-cell whose amplitudes either decay exponentially/grow exponentially/stay constant in time depending on the dissipation parameters. 
Following \cite{Zezyulin1018array,Mullerseaat6539}, (i) in general hold true 
when considering the  Kerr nonlinearity $g \mathcal{F}(\Psi_n ) \Psi_n $  in Eq.(\ref{eq:fb_eq}) 
where $\mathcal{F}( \Psi_n) = \sum_{j} |\psi_n^j|^{2} \ e_j\otimes e_j$ with \{$e_j\}$ canonical basis of $\mathbb{R}^\nu$. 
However, (ii) depends whether the linear CLS can be continued as compact time-periodic solution (breather) in the nonlinear regime of the lattice \cite{Danieli2018compact}.

%\section*{Main example}

\begin{figure}[!htbp]
\centering
\includegraphics [width=\columnwidth]{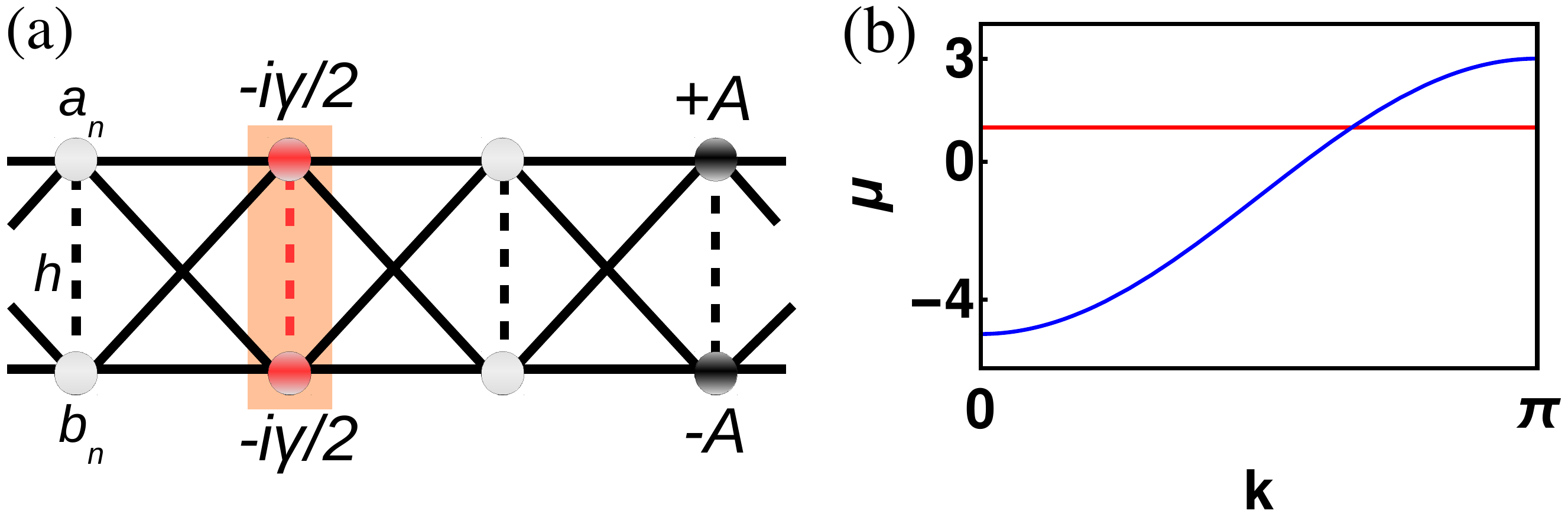}
\caption{(a) Profile of the
Cross-Stitch lattice. (b) Band structure for $h=1$.}
\label{figCS1}
\end{figure}

To illustrate these statements, we employ a simple one-dimensional two-bands network $\nu = 2$ with $\Psi_n = (a_n,b_n)^T$, as shown in Fig.\ref{figCS1}(a).
The matrices $H_0, H_1$  in Eq.(\ref{eq:fb_eq}) that define the Cross-Stitch lattice (CS) are
  \begin{equation}
  H_0 =  \begin{pmatrix}
       0 &h   \\[0.3em]
       h & 0    
        \end{pmatrix},
        \quad
          H_1 =  \begin{pmatrix}
       1 &1   \\[0.3em]
       1 & 1    
        \end{pmatrix},
        \quad
        V  = V_1 \equiv 
        \begin{pmatrix}
      1 & 0 \\[0.3em]
      0 & 1   
        \end{pmatrix}
  \label{eq:cs_matr}
  \end{equation}
For $\gamma = 0$, the linear CS network posseses
one dispersive band $\mu_1(k) = -h - 4\cos k$  and one flat band $\mu_2 = h$, as shown in Fig.\ref{figCS1}(b) for $h=1$.   

To fine-tune CPA in the CS network, 
in Eq.(\ref{eq:CS_ansatz}) we consider the normalized eigenvector $\varPhi_1 = \frac{1}{\sqrt{2}} (1,1)^T$ correspondent to the model's dispersive band $\mu_1(k)$.  
Considering the local symmetry of both the chain and  $\varPhi_1$, it naturally follows to choose a symmetric diagonal dissipative matrix $V = V_1$ in Eq.(\ref{eq:cs_matr}).  
This choice induce the CPA condition for a given $k$ - see Appendix \ref{sec:app1} for details. 
\begin{equation}
\gamma_* = 8\sin k
%\quad\Leftrightarrow\quad
%k_* = \arcsin \frac{\gamma}{8}
\label{eq:cpa_cs1}
\end{equation}  
Moreover, CPA occurs also in the nonlinear regime $g\neq 0$ of the lattice at the same condition Eq.(\ref{eq:cpa_cs1}) in perturbative regimes (small propagating waves regime) - as detailed in Appendix \ref{sec:app1}. 

\begin{figure}[!htbp]  
 \centering
\includegraphics[width=\columnwidth]{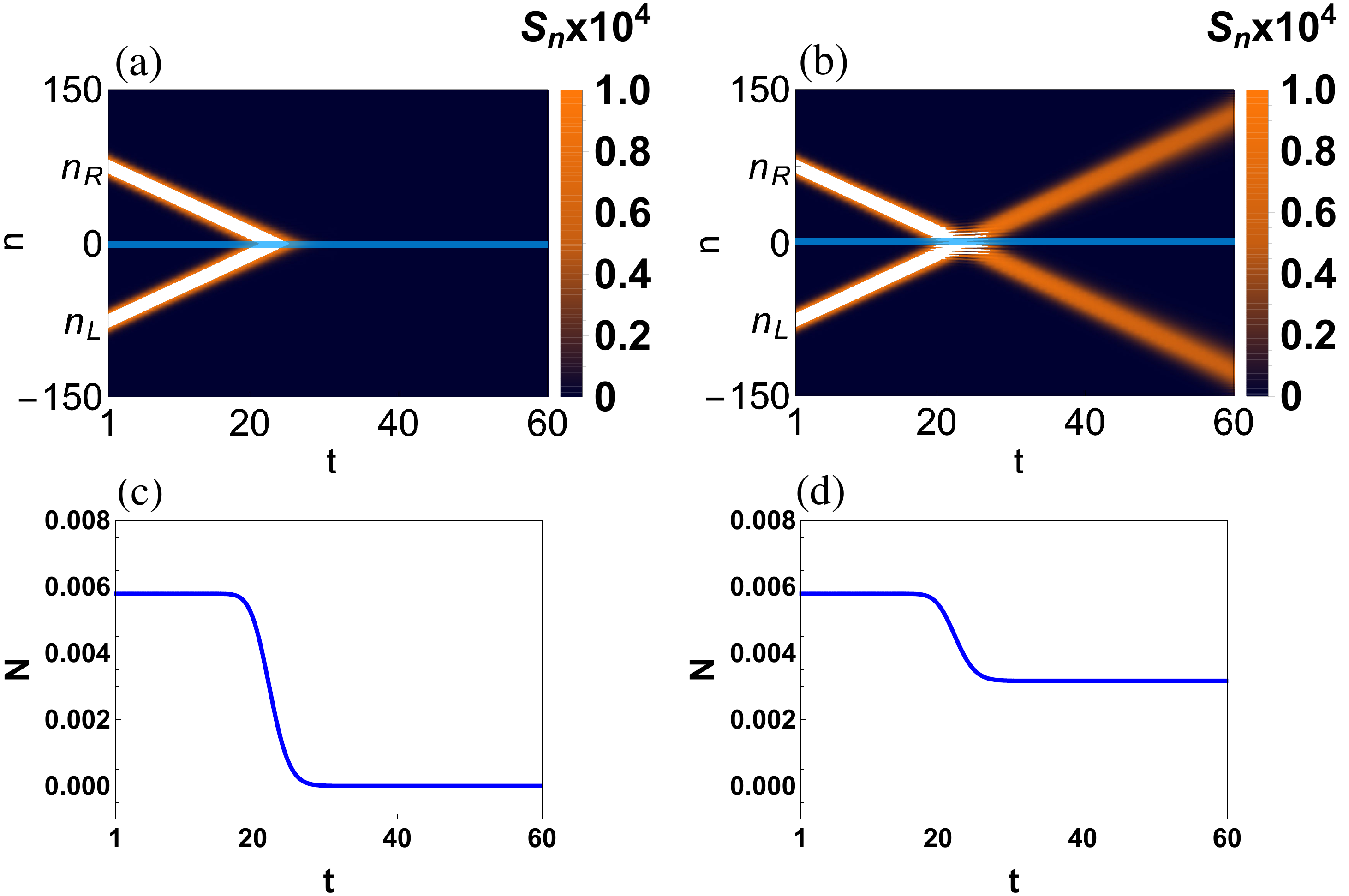}
 \caption{
Dispersive case $\Phi_1$: time evolution of $S_n$ (a) and $N$ (c) for $\gamma = 6.73 \approx \gamma_*$ and $S_n$ (b) and $N$ (d) for $\gamma = 1$.
Here: $k=1$ and $g=1$, while $n_R = 75$ and $n_L =-75$.  
The blue horizontal lines signal the dissipative unit-cell at $n=0$. 
}
  \label{fig:CS_lin_examples}
\end{figure}

We visualize numerically this theoretical prediction in the nonlinear CS network with nonlinear strength $g=1$ by considering two counter-propagating Gaussian beams \cite{Vicencio2007fano} centered far from the dissipative unit-cell $n=0$ at unit-cells $n_L\ll 0$ and $n_R \gg 0$ respectively, with $n_L = -n_R$. 
Written in components $\Psi_n = (a_n,b_n)^T$, the Gaussian beams at $t=0$ read 
\begin{equation}
\begin{split}
a_n = b_n &= P_0 e^{-\alpha (n - n_R)^2}e^{-ik(n - n_R)}\\
&+ P_0 e^{-\alpha (n - n_L)^2}e^{ik(n - n_L)} \\
\end{split}
\label{eq:CS_IC}
\end{equation}
where $P_0 = 0.01$ is the amplitude and $\alpha = 0.0075$ is the inverse width of the excitation 
- see Appendix \ref{sec:app2} for details. 
For $k=1$ in Eq.(\ref{eq:CS_IC}), the CPA condition Eq.(\ref{eq:cpa_cs1}) yields $\gamma_{*} \approx 6.73$.  
In Fig.\ref{fig:CS_lin_examples}(a,b) we show the time evolution of the local densities $S_n =  |a_n|^2 + |b_n|^2$ for $\gamma = \gamma_*\approx 6.73$ (a) and $\gamma = 1$  (b).
At the CPA condition - Fig.\ref{fig:CS_lin_examples}(a) - the two incoming radiations are fully absorbed at the dissipative cell $n=0$ (signaled with the blue horizontal line), and no propagating radiation follows.
Instead, away from the CPA condition - Fig.\ref{fig:CS_lin_examples}(b) - the two incoming radiations are only partially absorbed and propagating radiations survive. 
This complete and partial absorption is further visualized in Fig.\ref{fig:CS_lin_examples}(c) and (d) respectively, 
where we show the time-evolution of the total norm $N = \sum_n S_n$.

\begin{figure}[!htbp]  
 \centering
\includegraphics[width=\columnwidth]{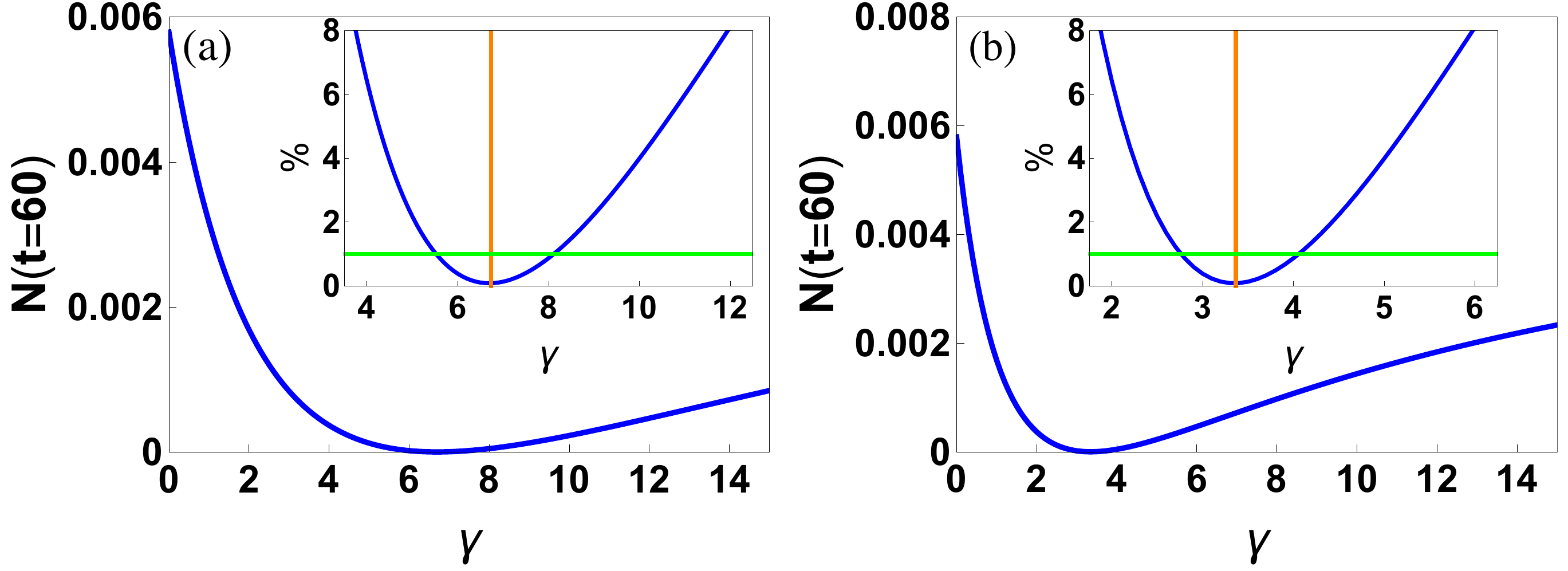}
 \caption{
Total norm $N$ at $t=60$ versus $\gamma$ for $V_1$ (a) and $V_2$ for $\delta = 1$(b). 
Inset: percentage \% at $t=60$, with $\gamma_*$ (orange);
$1\%$ of the norm (green). 
Here $h=1$ and $g=1$. 
}
  \label{fig:CS_lin_gamma_At}
\end{figure}

In Fig.\ref{fig:CS_lin_gamma_At}(a) we show the unabsorbed total norm $N$ at time $t=60$ computed for different values of the dissipation strength $\gamma$. 
This is further detailed in the inset, where we show the percentage of remaining norm  $\% = 100 \frac{N(t=60)}{N(t=0)}$ versus $\gamma$. 
The minimum of the curve lies to $\gamma_*$ - denoted by the vertical orange line - where $\%\approx 0$.
We can additionally observe that a significant part of the curve lies below the horizontal green line - line denoting one percent.
This indicates that there exists a non-negligible neighborhood of $\gamma_*$ where any $\gamma$ chosen within such interval leaves less than one percent of unabsorbed incoming radiations.

\begin{figure}[!htbp]  
 \centering
\includegraphics[width=\columnwidth]{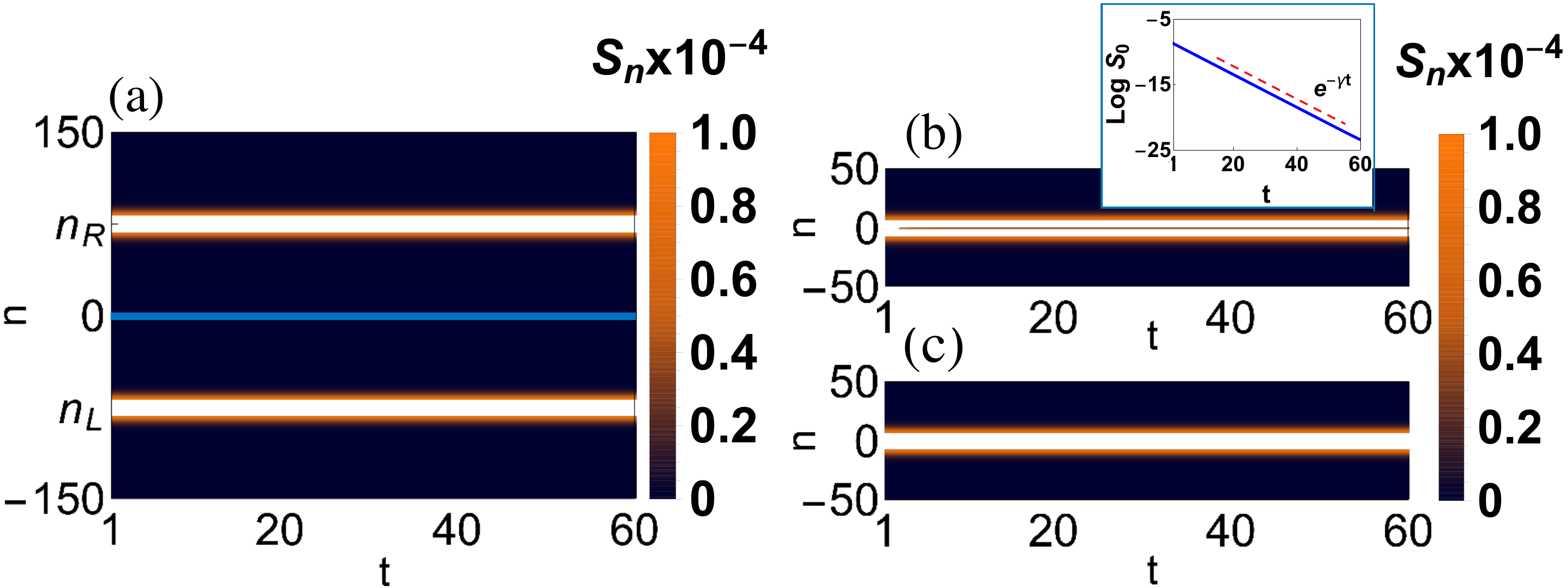}
 \caption{
Flat band case $\Phi_2$: 
(a) Time evolution of $S_n$ for $\gamma = 0.25$ and $\delta = 0$ while $n_R = 75$ and $n_L =-75$. 
(b) Same as in (a) with $n_R = 0 =n_L$. Inset: time evolution of $S_0$. %\sim e^{-\gamma t}$.
(c) Time evolution of $S_n$ for $\gamma = 0.25$, $\delta=1$ with $n_R = 0 =n_L$.  
Here $g=1$ and $k=1$. 
}
  \label{fig:CS_lin_CLS}
\end{figure}

Due to destructive interference,
the normalized eigenvector $\varPhi_2 = \frac{1}{\sqrt{2}} (1,-1)^T$ of $B(k)$ correspondent to the flat band $\mu_2$ of the CS lattice 
introduced in Eq.(\ref{eq:CS_ansatz}) does not yield propagating waves.
Indeed, considering $a_n = -b_n$ in Eq.(\ref{eq:CS_IC}) with $n_R\gg0$ and $n_L\ll 0$ yields two non-propagating Gaussian excitations centered at $n_R$ and $n_L$ respectively - as shown in Fig.\ref{fig:CS_lin_CLS}(a).   
This also persists for $n_L=0=n_R$  - as shown in Fig.\ref{fig:CS_lin_CLS}(b).   
However, the local density $S_n$ at $n=0$ of 
the non-propagating excitation decays exponentially in time as $S_0\sim e^{-\gamma t}$ - as shown in the inset of  Fig.\ref{fig:CS_lin_CLS}(b).

The emergence of the dissipative non-propagating excitation at $n=0$ shown in Fig.\ref{fig:CS_lin_CLS}(b) follows from the fact that the dissipation $V_1$ in Eq.(\ref{eq:cs_matr})  is symmetric between the two sub-lattices and therefore it does not break the CLS located at the unit-cell $n=0$ -  state indicated with black dots in Fig.\ref{figCS1}(a) - but it turns its frequency complex,  $\mu = h-i\gamma/2$.
Hence, this CLS turns dissipative. 
Analogous fate occurs in the nonlinear regime of the CS lattice. 
Indeed, as discussed in Ref.\cite{Danieli2018compact}, Kerr nonlinearity preserves destructive interference, and the dissipative CLS at $n=0$ with amplitude $A$  
\begin{equation}
U_{n,0}(t) = A
\begin{pmatrix}
      1 \\[0.3em]
      -1   
\end{pmatrix}
\delta_{n,0}
e^{-i\Omega t}
\label{eq:CB_CS1}
\end{equation}
becomes a dissipative solution of the lattice with renormalized frequency $\Omega = h + g A^2 - i\gamma/2$.  % -  state dubbed {\it dissipative compact breather}.}
In both linear and nonlinear regime of the CS, 
the local density of the compact solution decays exponentially in time 
$S_0(t) \sim e^{-\gamma t}$ due to the imaginary term $- i\gamma/2$ in their frequency.  
This behavior in time 
can be controlled by employing a second dissipation matrix $V_2$ in Eq.(\ref{eq:cs_matr}) for $\delta\in \mathbb{R}$
\begin{equation}
    V=    V_2 \equiv 
        \begin{pmatrix}
      1 & \delta \\[0.3em]
      \delta & 1   
        \end{pmatrix}
  \label{eq:cs_diss_2}
  \end{equation}
The frequency of the compact dissipative solution Eq.(\ref{eq:CB_CS1}) reads $\Omega = h + g A^2 - i \gamma (1 - \delta)/2$.
Hence $S_0(t) \sim e^{ - \gamma  (1 - \delta) t}$, which recalling that for any $\gamma>0$ implies that the local density $S_0$ of the compact solution decays exponentially for $\delta<1$, grows exponentially for $\delta>1$ and stay constant for $\delta = 1$. 
This consequently implies the existence of a non-propagating excitations located at the dissipative unit-cell
% which emulates the laser-absorber set-up \cite{Longhi:2010,Chong:2011PT,Mostafazadeh_2012}, and 
with the amplitude at the dissipative cell  $n=0$ that either decays, grows or stays constant with time according to the dissipation parameters  - as shown in Fig.\ref{fig:CS_lin_CLS}(c) for $\delta = 1$.  

This second dissipative matrix $V_2$ in Eq.(\ref{eq:cs_diss_2}) not only allows for fine control of the time-behavior of the compact flat band solutions, but it also induces CPA with condition 
\begin{equation}
\gamma_* = \frac{8}{1+\delta}\sin k
%\quad\Leftrightarrow\quad
%k_* = \arcsin \frac{\gamma(1+\delta)}{8}
\label{eq:cpa_cs2}
\end{equation}  
that generalizes Eq.(\ref{eq:cpa_cs1}) - see Appendix \ref{sec:app1}  for details. 
This prediction has been numerically confirmed in Fig.\ref{fig:CS_lin_gamma_At}(b) in the same framework as for $V_1$.
In particular, we can again observe in the inset that while CPA is achieved at $\gamma_*\approx 3.115$ - condition obtained from Eq.(\ref{eq:cpa_cs2}) with $k=1$ and $\delta = 1$ - a non-trivial neighborhood of $\gamma_*$ exists where less than one percent of the incoming beams' total norm is unabsorbed.

%\section*{Generalization and Conclusion}

\begin{figure}[!htbp]
\centering
\includegraphics [width=0.95\columnwidth]{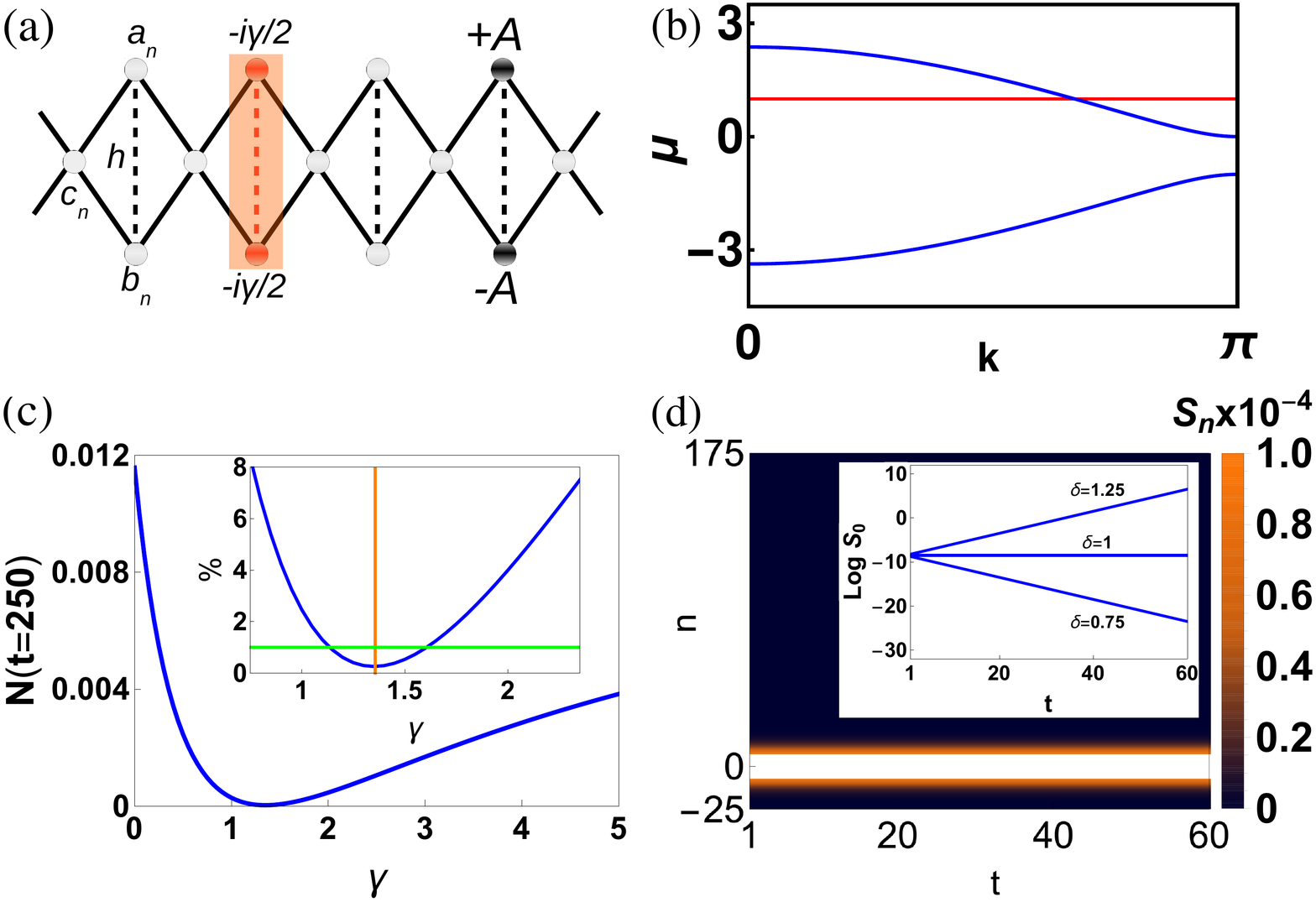}
\caption{(a) Profile of the
diamond chain. (b) Band structure.
(c) Total norm $N$ at  $t=250$ versus $\gamma$  for $\delta=1$. 
Inset: percentage \% at $t=250$, with $\gamma_*$ (orange);
$1\%$ of the norm (green). 
(d) Time evolution of $S_n$ for $\gamma = 1$, $\delta = 1$. 
Inset: time evolution of $S_0$  for $\delta=1.25$ (top), $\delta=1$(middle) and $\delta = 0.75$ (bottom) with $\gamma = 1$. 
Here $g=1$ and $h=1$.
}
\label{fig:diam}
\end{figure}

Remarkably, this procedure works and CPA can be obtained in a plethora of other flat band topologies. 
For example, the diamond chain, whose profile is shown in Fig.\ref{fig:diam}(a) and the equations of motion written in components $\Psi_n=(a_n,b_n,c_n)^T$ with local dissipation controlled by the parameters $\gamma,\delta$ and in presence of Kerr nonlinearity are shown below 
\begin{equation}
\begin{split}
i\dot{a}_n &= - c_{n}  - c_{n+1} -  h   b_{n}  -  \frac{i\gamma}{2} (a_0 + \delta b_0)\delta_{n,0} + g a_n|a_n|^2\\
i\dot{b}_n &= - c_{n}  - c_{n+1} -  h   a_{n}  -  \frac{i\gamma}{2} (b_0 + \delta a_0)\delta_{n,0} + g b_n|b_n|^2\\
i\dot{c}_n &= - a_{n-1} - a_{n} - b_{n-1} - b_{n}   + g a_n|a_n|^2
\end{split}
\label{eq:DC_nl_V1}
\end{equation}
For $g=0$ and $\gamma=0$, Eq.(\ref{eq:DC_nl_V1}) possesses two dispersive bands $\mu_{1,2}(k) = (- h \pm \sqrt{h^2+16+16 \cos k})/2$ and one flat band $\mu_3=h$  - Fig.\ref{fig:diam}(b). 
This model (also called rhombic lattice) has been employed to theoretically study the impact of non-hermitian hopping \cite{Leykam2017flatbandNH}, magnetic field \cite{Vidal2000interaction,Diliberto2018nonlinear,Grigoric1029nonlinear}  and electric fields \cite{Khomeriki2016landau} on flat band networks,  
%Further, this system  %and its similarly-shaped versions (saw-tooth lattice) 
%have been employed 
as well as to experimentally realize compact localized states \cite{Mukherjee2015Observation2}, 
%realize compact localization 
study CLS in driven photonic flatband lattices \cite{Mukherjee2017Observation}, 
and experimentally observe Aharonov-Bohm caging effect \cite{Mukherjee2018experimental}. 
% and experimentally realize compact states in $\mathcal{PT}$-symmetric rhombic lattice \cite{Biesenthal2019experimental}. }

Via the transfer matrix formalism, 
CPA can be fine-tuned in the diamond chain considering the eigenmodes $\phi_{1,2} = (1,1,\pm \sqrt{2} e^{-k/2})^T / 2$ in the ansatz 
Eq.(\ref{eq:CS_ansatz}) correspondent to the dispersive bands $\mu_{1,2}$.
For both bands, CPA occurs with condition 
\begin{equation}
\gamma_* = \frac{4\sqrt{2}}{1+\delta}\sin \frac{k}{2}
%\quad\Leftrightarrow\quad
%k_* = 2 \arcsin \frac{\gamma}{4\sqrt{2}}
\label{eq:cpa_dc1}
\end{equation}  
condition which still holds 
in the nonlinear regime $g\neq 0$ of the diamond chain Eq.(\ref{eq:DC_nl_V1}).

This theoretical prediction can be achieved with two counter-propagating Gaussian beams located at $n_L\ll 0$ and $n_R\gg 0$ respectively with $n_L = -n_R$, which in the diamond chain case are  
$b_n = a_n =  P_0 e^{-\alpha (n - n_R)^2}e^{-ik(n - n_R)} + P_0 e^{-\alpha (n - n_L)^2}e^{ik(n - n_L)}$ 
and $c_n = \pm \sqrt{2} e^{-k/2} a_n$ - see Appendix \ref{sec:app2}  for details.  
For $k=1$ and $\delta = 1$, Eq.(\ref{eq:cpa_dc1}) predicts that CPA is achieved for $\gamma_* \approx 1.36$, which is visualized analogously with the CS in Fig.\ref{fig:diam}(c). 
On the other hand, the flat band eigenvector $\varPhi_3 = \frac{1}{\sqrt{2}} (1,-1,0)^T$ of $B(k)$ of the diamond chain induces a non-propagating Gaussian excitation located at $n=0$  -  Fig.\ref{fig:diam}(d). 
The reason for this phenomenon follows from the fact that the chosen dissipation does not lift the linear CLS at $n=0$ - indicated with black dots in Fig.\ref{fig:diam}(a) -  but it only alters its frequency and makes it dissipative. 
This still holds in the nonlinear regime $g\neq 0$ of Eq.(\ref{eq:DC_nl_V1}) since the linear CLS locate at the unit-cell $n=0$ can be continued as compact breather with frequency  
$\Omega = h + g A^2 - i\gamma(1 - \delta)/2$ \cite{Danieli2018compact}.  
This yields a precise control of the time behavior of the breather density $S_0$, namely
 $S_0$ decays exponentially for $\delta<1$, grows exponentially for $\delta>1$ and stay constant for $\delta = 1$  -  as shown in the inset of Fig.\ref{fig:diam}(d).

By choosing certain flat band geometries, we showed that local dissipations can be arranged to induce CPA and at the same time preserve the compact solution that exists at the dissipative unit-cell.
This notably relates to the local symmetry in the chosen dissipative terms (matrix $V$) as shown in Appendix \ref{sec:app1}.  
If this symmetry in $V$ is broken, the destructive interference that yields CLS is locally lifted and the system does not admit dissipative compact states - although it still can support CPA. 
%However, such asymmetric dissipative unit-cell still can supports CPA. 
The current setting of a single dissipative unit-cell in Eq.(\ref{eq:fb_eq}) prevent to simultaneously achieve CPA and dissipative CLS in networks whose compact states occupies at least $U\geq 2$ unit-cells -  {\it e.g.} the Lieb lattice \cite{Nita2013spectral} - since the partial overlap between CLS and dissipation lifts destructive interference. 
Hence, the existence of dissipative CLS requires the local dissipation to span over at least $U$ unit-cells with a profile which respect the CLS symmetries.
Such dissipative CLS may also become dissipative compact solutions of the nonlinear regime - in agreement with Ref.\cite{Danieli2018compact} - as well as their time behavior (exponential growth, decay or being constant) can be controlled by tuning the dissipation parameters, in analogy with what we herewith reported. 
Moreover, following \cite{Zezyulin1018array,Mullerseaat6539}, we observe that the absorption conditions in Eq.(\ref{eq:cpa_cs1},\ref{eq:cpa_cs2},\ref{eq:cpa_dc1}) can be reversed into lasing conditions by permuting the sign of the prefactor $i\gamma/2$ 
of the dissipative terms $V$ in Eq.(\ref{eq:fb_eq}) to positive $+i\gamma/2$. 
This namely implies that in Eq.(\ref{eq:CS_ansatz}) % - and visualized in  Fig.\ref{fig:CPA} - 
the incoming radiations $a_L, b_R$ are annihilated and only outgoing radiations $a_R, b_L$ emerge from the 
non-hermitian unit-cell.

To summarize, we have shown that local dissipations in lattice networks can yield simultaneously two wave phenomena: coherent perfect absorption and the existence of dissipative compact localized states.  Both phenomena are the result of destructive wave interference, and we shown that they can be embedded in a single device which posses both dispersive and flat bands by engineering the local dissipations. 
In a broader perspective, our results firstly usher the existence and the study of compact dissipative breathers in nonlinear non-Hermitian media (see \cite{Eckmann2019decay} for a recent survey on dissipative discrete breathers) which may additionally account for Ghost states in the case of $\mathcal{PT}$-symmetric nonlinear networks \cite{rodrigues2012pt}. 
Additionally, it is interesting to note the analogy between the dissipative compact solutions induced by symmetric local dissipation with the bound states in the continuum (BIC) waves studied in Ref.\cite{Hsu2016bound}. 
Secondly, these findings pave the way to accomplish CPA and lasing phenomena in
multi band networks supporting propagating waves. 
These include optical and photonic systems as well as exciton-polaritons and microwave systems, all systems where flat band networks have been studied and experimentally realized.
Moreover the Gaussian beams set-up herewith employed to visualize our theoretical findings offers a novel and powerful scheme to experimentally realize CPA in order to
fabricate switches \cite{Fang2014fang}, interferometers \cite{Wan2011time} and logic elements \cite{Papaioannou2016all} 
in several physically relevant frameworks, particularly nonlinear optical and photonic lattice networks.
\\

% It is shown that such complex eigen value accounts for the Ghost states in $\mathcal{PT}-$ symmetric systems 
%\cite{rodrigues2012pt}.
%\cite{Eckmann2019decay}

The  authors  acknowledge  financial  support  from the Institute for Basic Science (IBS)-Project Code No. IBS-R024-D1. 
We thank S. Flach, A. Andreanov and P. Kevrekidis for helpful discussions.

\appendix

\section{Coherent perfect absorption conditions} \label{sec:app1}

 \subsection{Transfer Matrix Method}

Let us recap the ansatz %Eq.(\ref{eq:CS_ansatz}) 
  \begin{equation}
   \begin{split}
    U_n & = \Phi_j\left\{
                \begin{array}{ll}
                  \alpha_L e^{ikn}+\beta_L e^{-ikn} ~~~n < 0 \\
                  u_0 ~~~~~~~~~~~~~~~~~~~~~~~n=0  \\
                 \alpha_R e^{ikn}+\beta_R e^{-ikn} ~~~n > 0.
                \end{array}
              \right. 
  \label{eq:CS_ansatz_sm}
     \end{split}
  \end{equation}
For  $\phi_j$ the eigenmode of the matrix Bloch $B(k)$ correspond to one dispersive band $\mu_j$ of a multiband  network.
We define 
\begin{equation}
 a_L = \alpha_L\Phi_j\ ,\quad  a_R = \alpha_R\Phi_j \ ,\quad  b_L = \beta_L\Phi_j \ ,\quad  b_R = \beta_R\Phi_j 
 \label{eq:CS_ansatz_sm_2}
\end{equation}  
where $a_L$ and $b_R$ are the incident wave amplitudes from left ($L$) and right ($R$), while $a_R$ and $b_L$ are the reflected wave amplitudes. 
We then introduce the transfer matrix $M$, defined as 
  \begin{equation}
   \begin{pmatrix}
       a_R  \\[0.3em]
       b_R   
        \end{pmatrix} 
   = M
 \begin{pmatrix}
       a_L  \\[0.3em]
       b_L    
        \end{pmatrix}
        \quad \text{where}\quad 
        M =  \begin{pmatrix}
       M_{11}(k)  & M_{12}(k) \\[0.3em]
       M_{21}(k) &  M_{22}(k)   
        \end{pmatrix} 
       \label{eq:TM_sm}
  \end{equation}
 In this case, the transfer matrix  $M$ is a square matrix of size $2\nu$ and the components $M_{i,j}(k)$ blocks of side $\nu$. \\
 \\
Following \cite{Mullerseaat6539}, CPA conditions can be derived via the transfer matrix $M$. 
CPA occurs when there exist non-zero incoming radiations $a_L$ and $b_R$, but the outgoing radiations $a_R$ and $b_L$ are zero.
In Eq.(\ref{eq:TM_sm}), this translates into finding the $k_{\ast}$ such that $M_{11}(k_{\ast}) = \mathbb{O}_\nu$ \cite{Zezyulin1018array,Mullerseaat6539}, since 
   \begin{equation}
   \begin{pmatrix}
       0  \\[0.3em]
       b_R   
        \end{pmatrix} 
        = \begin{pmatrix}
    M_{11}(k_{\ast})  a_L   \\[0.3em]
     M_{21}(k_{\ast}) a_L  
        \end{pmatrix},
        \quad\Leftrightarrow\quad
        \begin{cases}
        M_{11}(k_{\ast}) = \mathbb{O}_\nu\\
        b_R =  a_L 
        \end{cases}
       \label{eq:cpatm2}
  \end{equation}  
and $M_{21}(k_{\ast}) = 1$. Here, $(k_{\ast})^2$ represents a time-reversed spectral singularity.
Let us remark that reversing the condition -  incoming radiations $a_L$ and $b_R$ are zero, but the outgoing radiations $a_R$ and $b_L$ are non-zero - yields lasing condition \cite{Chong:2011PT}. 
In Eq.(\ref{eq:TM_sm}), this translates into finding the $k_{\ast}$ such that $M_{22}(k_{\ast}) = \mathbb{O}_\nu$ \cite{Zezyulin1018array,Mullerseaat6539}.

\subsection{Cross-Stitch - Linear Regime}\label{sec:cross_stitch_lin}

 The cross-stitch lattice is a $\nu=2$ bands problem, 
 \begin{equation}
 \begin{split}
\mu U_n = & - H_0 U_n -  H_1 U_{n+1} - H_1^\dagger U_{n-1}-i\frac{\gamma}{2} V U_0 \delta_{n,0}\ ; \qquad \qquad 
H_0 =  \begin{pmatrix}
       0 &h   \\[0.3em]
       h & 0    
        \end{pmatrix},
        \qquad
          H_1 =  \begin{pmatrix}
       1 &1   \\[0.3em]
       1 & 1    
        \end{pmatrix}
\end{split}
\label{eq:fb_eq2_sm}
\end{equation}
with one dispersive band $\mu_1(k) = -h - 4\cos k$, whose correspondent eigenvector is $\Phi_1 = \frac{1}{\sqrt{2}} (1,1)^T$.  
In this case, we choose two type of dissipations matrix $V$  defined as
\begin{equation}
V_1 = 
        \begin{pmatrix}
      1 & 0 \\[0.3em]
      0 & 1   
        \end{pmatrix} 
\qquad and \qquad 
V_2 = 
        \begin{pmatrix}
      1 & \delta \\[0.3em]
      \delta & 1   
        \end{pmatrix}
\label{eq:V_cs_fb}
\end{equation}

\subsubsection*{Case 1 - $V_1$}\label{sec:cs_diss1}

For the Cross-Stitch lattice $H_1 = H_1^\dagger $. 
Eq.(\ref{eq:fb_eq2_sm}) at $n=\{0, \pm1\}$ yield
      \begin{equation}
   \begin{split}
&\qquad  \left[\mu_1 \mathbb{I}_2 + H_0 + i\frac{\gamma}{2} V_1 \right] U_0 \\
& = - H_1 \left[ a_R e^{i k }+b_R e^{-i k } + a_L e^{-i k }+b_L e^{i k } \right]
  \label{eq:CS_v1_1_sm}
     \end{split}
  \end{equation}
         \begin{equation}
   \begin{split}
 \left[\mu_1 \mathbb{I}_2 + H_0 \right]  \left[ U_1+ U_{-1} \right]=  
 -  H_1 (U_{2}  +  U_{-2})  - 2 H_1  U_0
  \label{eq:CS_v1_2_sm}
     \end{split}
  \end{equation}
      \begin{equation}
   \begin{split}
 \left[\mu_1 \mathbb{I}_2 + H_0 \right]  \left[ U_1 - U_{-1}  \right] = 
  -  H_1 (U_{2}  -  U_{-2})  
  \label{eq:CS_v1_3_sm}
     \end{split}
  \end{equation}
From Eq.(\ref{eq:CS_ansatz_sm_2}), we sort Eq.(\ref{eq:CS_v1_1_sm},\ref{eq:CS_v1_2_sm},\ref{eq:CS_v1_3_sm}) in terms of $\Phi_1$.
Recalling the identities
$H_0\Phi_1 = h \Phi_1$, $H_1\Phi_1 = 2 \Phi_1$, and $ [ \mu_1 \mathbb{I}_2 + H_0 ]\Phi_1  = -4\cos k \Phi_1$, Eq.(\ref{eq:CS_v1_2_sm}) and Eq.(\ref{eq:CS_v1_3_sm}) reduce to
         \begin{equation}
   \begin{split}
[ \alpha_R +  \beta_R + \alpha_L   +  \beta_L   ]  \Phi_1 =  H_1  U_0 
  \label{eq:CS_v1_4_sm}
     \end{split}
  \end{equation}
      \begin{equation}
   \begin{split}
[ \alpha_R +  \beta_R ]\Phi_1 = [ \alpha_L   +  \beta_L   ]  \Phi_1 \\
  \label{eq:CS_v1_5_sm}
     \end{split}
  \end{equation}
Since the identity
$ \left[\mu_1 \mathbb{I}_2 + H_0 + i\frac{\gamma}{2} V_1 \right] \Phi_1 = \left(- 4\cos k +\frac{i\gamma}{2} \right) \Phi_1$, 
from Eq.(\ref{eq:CS_v1_1_sm}) it follows 
      \begin{equation}
   \begin{split}
U_0 
  &=  - \frac{2}{-4\cos k + i\frac{\gamma}{2} } [ \alpha_R e^{i k }+ \beta_R e^{-i k } + \alpha_L e^{-i k }+ \beta_L e^{i k } ] \Phi_1
%  &=  - 2 \left(-4\cos k + i\frac{\gamma}{2} \right)^{-1} [ \alpha_R e^{i k }+ \beta_R e^{-i k } + \alpha_L e^{-i k }+ \beta_L e^{i k } ] \Phi_1
  \label{eq:CS_v1_6_sm}
     \end{split}
  \end{equation}
  In Eq.(\ref{eq:CS_v1_4_sm}), this ultimately results into the equality 
          \begin{equation}
   \begin{split}
&\quad\left [ \alpha_R \left(4\sin k  + \frac{\gamma}{2} \right)   -  \beta_R   \left(4\sin k   - i\frac{\gamma}{2} \right)   \right]  \Phi_1 \\ 
& = \left[  \alpha_L   \left( 4 \sin k   - \frac{\gamma}{2} \right)   -  \beta_L  \left(4i \sin k +  i\frac{\gamma}{2} \right)  \right]  \Phi_1
  \label{eq:CS_v1_7_sm}
     \end{split}
  \end{equation} 
Since $\Phi_1 =\frac{1}{\sqrt{2}} (1,1)^T$, we just refer at both Eq.(\ref{eq:CS_v1_5_sm}) and Eq.(\ref{eq:CS_v1_7_sm}) as equations of scalar numbers. 
These two conditions
 ultimately yield the entrees of the transfer matrix $M$ Eq.\eqref{eq:cpatm2}
    \begin{equation}
     \begin{split}
   M_{11} &=  \frac{8 \sin k-\gamma}{8 \sin k} \mathbb{I}_2~~~~M_{12}=\frac{- \gamma}{8 \sin k} \mathbb{I}_2 \\
   M_{21} &=   \frac{\gamma}{8 \sin k}\mathbb{I}_2 \qquad \quad M_{22}=\frac{8 \sin k+\gamma}{8 \sin k}\mathbb{I}_2
    \label{eq:CS_v1_8_sm}
     \end{split}
  \end{equation}
Imposing $ M_{11} (k) = \mathbb{O}_2$ yields the CPA condition 
\begin{equation}
\gamma_* = 8\sin k
\quad\Leftrightarrow\quad
k_* = \arcsin \frac{\gamma}{8}
\label{eq:CS_v1_9_sm}
\end{equation}

\subsubsection*{Symmetry imposed coexistence of CLS and CPA}

\begin{figure}[H]  
 \centering
\includegraphics[width=0.95\columnwidth]{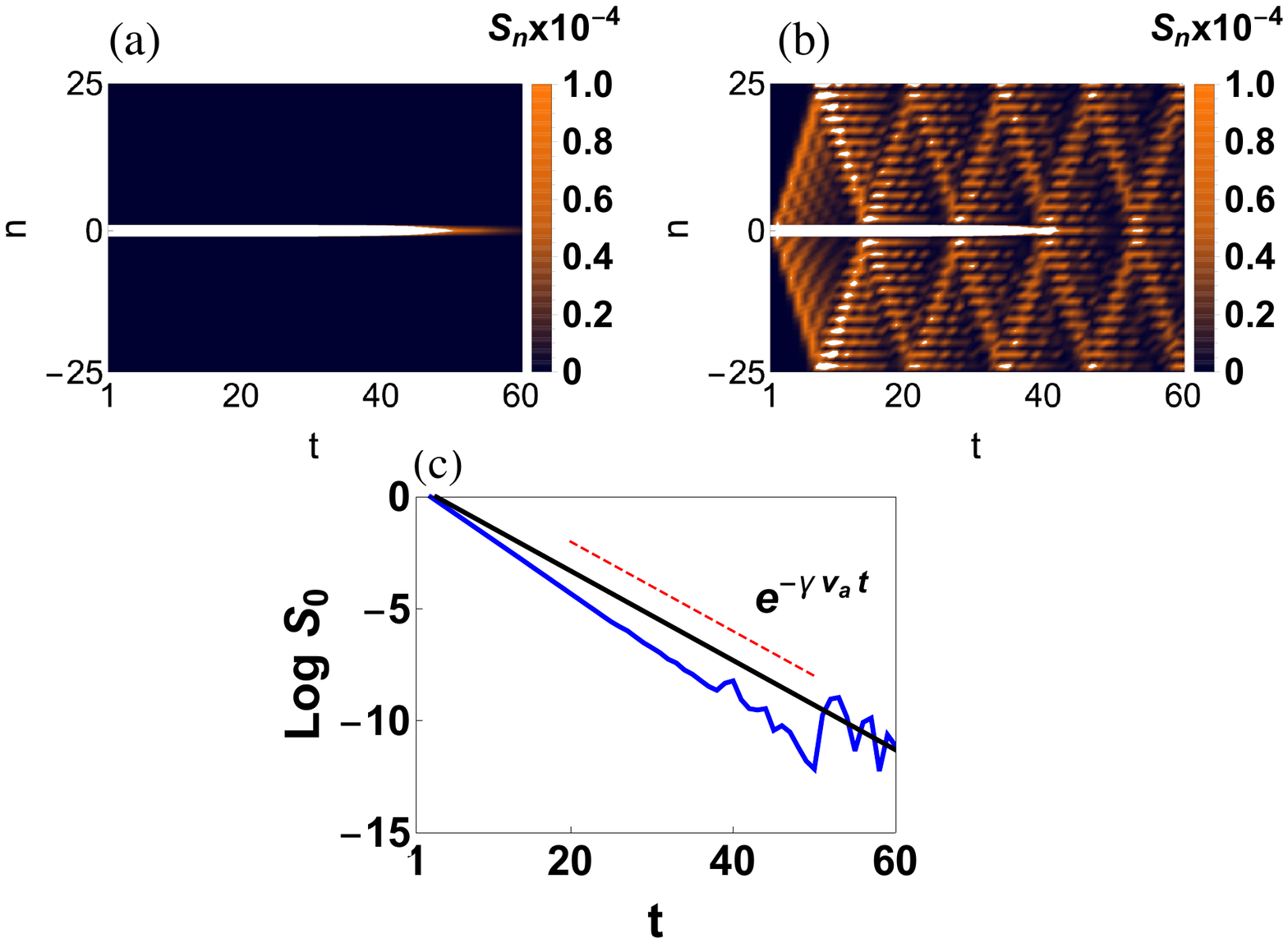}
 \caption{
(a) Time evolution of $S_n$ for $v_a = 1 = v_b$. 
(b) Time evolution of $S_n$ for  $v_a = 1$ and $v_b=1.25$. 
(c) Time evolution of $S_0$ for $v_a = 1 = v_b$ (black) and   for $v_a = 1$ and $v_b=1.25$ (blue). The red dashed line indicates the exponential decay $S_0 \sim e^{-\gamma v_a t}$. }
  \label{fig:app0}
\end{figure}

\begin{figure}[H]  
 \centering
\includegraphics[width=0.6\columnwidth]{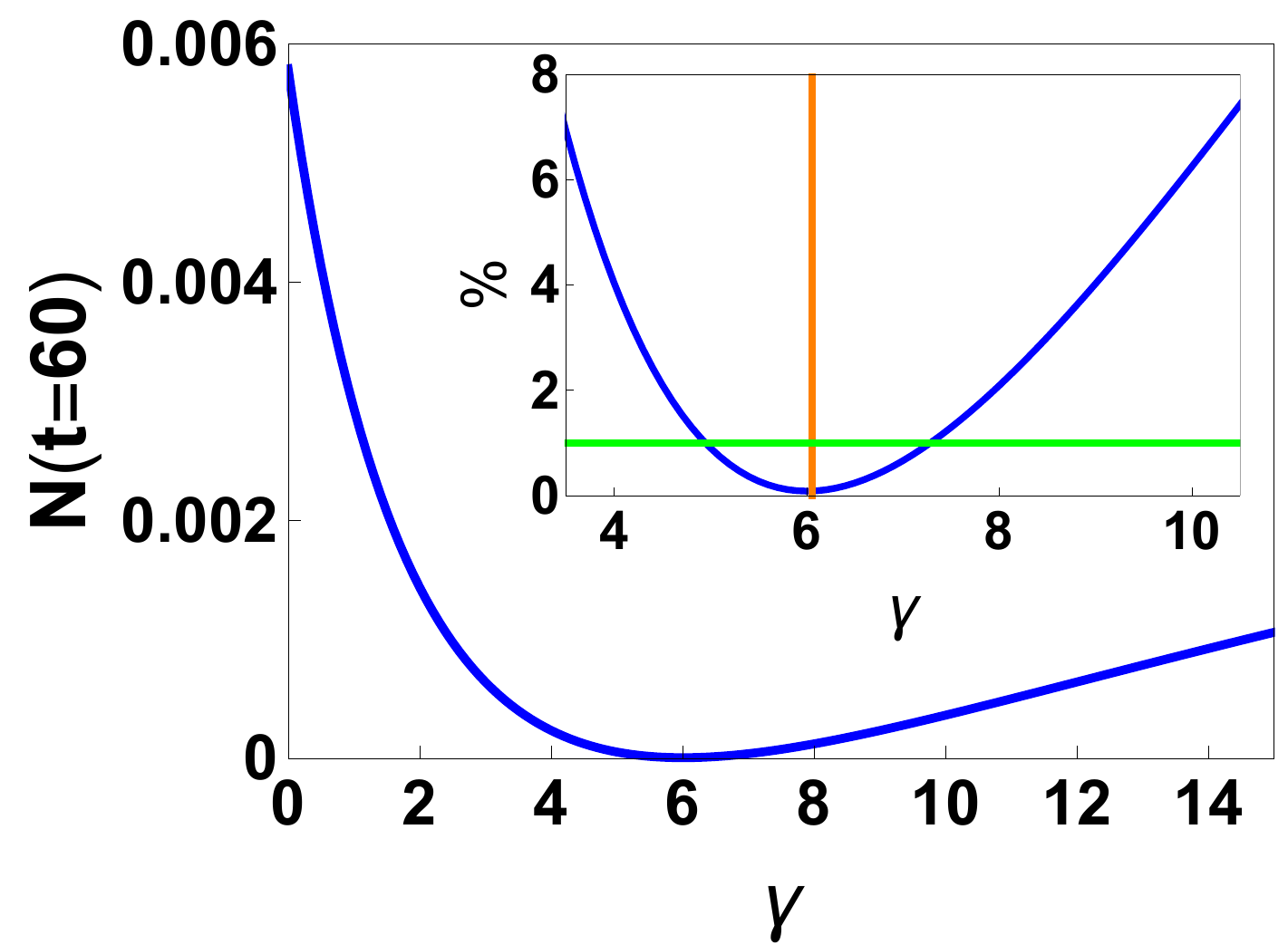}
 \caption{
 Total norm $N$ at $t=60$ versus $\gamma$ for $v_a = 1$ and $v_b=1.25$. 
Inset: percentage \% at $t=60$, with $\gamma_*$ (orange);
$1\%$ of the norm (green). }
  \label{fig:app0b}
\end{figure}

Let us consider the Cross-Stitch lattice (CS) lattice with diagonal elements $v_a,v_b$ in the matrix $V_1$ in Eq.(\ref{eq:V_cs_fb}) - Eq.(4) of the main text
\begin{equation}
\begin{split}
\mu a_n &= - a_{n-1} - a_{n+1} - b_{n-1} - b_{n+1} - h b_{n}  - \frac{i \gamma}{2}  v_a a_0 \delta_{n,0} \\
\mu b_n &= - b_{n-1} - b_{n+1} - a_{n-1} - a_{n+1} - h a_{n}  - \frac{i\gamma}{2}  v_b  b_0 \delta_{n,0}.
\end{split}
\label{eq:CS_S1}
\end{equation}
For $v_a = v_b$, the case in the former section is obtained, which leads to CPA as well as to dissipative CLS with frequency $\Omega = h - i\gamma / 2$. 
For $v_a \neq v_b$, the two components of the compact state $a_0,b_0$ decay at different rate, namely $|a_0|^2\sim e^{-\gamma v_a t}$ and  $|b_0|^2\sim e^{-\gamma v_b t}$.
This lifts destructive interference and consequently destroy the compact state.
This is shown in Fig.\ref{fig:app0}, where the time-evolution of $S_n$ is shown for the CLS with $\gamma = 0.2$ and $v_a = 1 = v_b$ in (a) and $v_a = 1$ while $v_b=1.25$ in (b), confirming that in the former case compactness is preserved and in the latter case compactness is lost.
In panel (c) we show how the exponential decay of $S_0\sim e^{-\gamma v_a t} $ for $v_a = v_b$ (black curve) is distorted and lost  for $v_a \neq v_b$ (blue curve).  
\noindent
However, such asymmetric dissipation does not lift CPA for $v_a\neq v_b$ but $v_a - v_b \approx 0$, and the CPA. Indeed, condition Eq.(5) of the main text gains a prefactor
\begin{equation}
\gamma_* =  \frac{v_a + v_b}{2v_a v_b} 8\sin k
\label{eq:CPA_asym}
\end{equation}
as shown in Fig.\ref{fig:app0b}

\noindent
This can be understood by employing a unitary transformation as in Ref.\cite{Flach2014detangling}
\begin{equation}
\begin{split}
p_n &= \frac{1}{\sqrt{2}}(a_n+b_n),~~~q_n= \frac{1}{\sqrt{2}}(a_n-b_n)\\
v_{+} &= \frac{1}{2}(v_a+v_b),~~~~~~~v_{-}= \frac{1}{2}(v_a-v_b),
\end{split}
\label{eq:CS_S2}
\end{equation}
which maps Eq.~\eqref{eq:CS_S1} to
\begin{equation}
\begin{split}
\mu p_n &=  \frac{i \gamma}{2} v_{+} \delta_{n,0}p_0 - h p_n+\frac{i \gamma}{2}  v_{-}\delta_{n,0}q_0  -2(p_{n+1}+p_{n-1}) \\
\mu q_n &=  \frac{i \gamma}{2} v_{+} \delta_{n,0} q_0 +h q_n +\frac{i \gamma}{2} v_{-}\delta_{n,0}p_0.
\end{split}
\label{eq:CS_S2}
\end{equation}
For $v_a=v_b$, then $v^{-}=0$ which fully decoupled the two sub-lattices. The $p_n$ sub-lattice in Eq.(\ref{eq:CS_S2}) yields the CPA condition $\gamma_* = 8\sin k$ in Eq.(\ref{eq:CS_v1_9_sm}). 
For $v_a\neq v_b$ with $v_a - v_b \approx 0$ will not largely renormalize $\mu$ in the $q_n$ sub-lattice, hence at $n=0$ we approximate  $\mu\approx h$. It then follows that $q_0 = - \frac{v^-}{v^+} p_0$ which substituted in the $p_n$ sub-lattice yields
\begin{equation}
\begin{split}
\mu p_n &=  \frac{i \gamma}{2} \left( v_{+}  - \frac{v_{-}^2}{v_+}\right)  p_0 \delta_{n,0}  - h p_n -2(p_{n+1}+p_{n-1}) \\
\end{split}
\label{eq:CS_S3}
\end{equation}
The prefactor of the dissipative strength $\gamma$ between brackets can be recast into
\begin{equation}
\begin{split}
  \frac{v_{+}^2 -  v_{-}^2}{v_+} &= \frac{2}{ v_a+v_b} \frac{v_a^2+v_b^2 + 2v_a v_b -(v_a^2+v_b^2 - 2v_a v_b) }{4}  \\
&  = \frac{2 v_a v_b}{ v_a+v_b} 
\end{split}
\label{eq:CS_S4}
\end{equation}
and it ultimately enters the CPA condition in Eq.(\ref{eq:CPA_asym}).

\subsubsection*{Case 2 - $V_2$}\label{sec:cs_diss2}

The second case follows the former one, although with a different identity
$ \left[\mu_1 \mathbb{I}_2 + H_0 + i\frac{\gamma}{2} V_2 \right] \Phi_1 = \left[- 4\cos k +\frac{i\gamma}{2}(1+\delta) \right] \Phi_1$
employed in Eq.(\ref{eq:CS_v1_6_sm}). 
It then follows 
      \begin{equation}
   \begin{split}
U_0 
  = & - \frac{2}{-4\cos k + i\frac{\gamma}{2} (1+\delta)} \cdot \\
  &\quad \cdot [ \alpha_R e^{i k }+ \beta_R e^{-i k } + \alpha_L e^{-i k }+ \beta_L e^{i k } ] \Phi_1
%  &=  - 2 \left[-4\cos k + i\frac{\gamma}{2} (1+\delta) \right]^{-1} [ \alpha_R e^{i k }+ \beta_R e^{-i k } + \alpha_L e^{-i k }+ \beta_L e^{i k } ] \Phi_1
  \label{eq:CS_v2_1_sm}
     \end{split}
  \end{equation}
Similar procedure lead to the transfer matrix entrees
    \begin{equation}
     \begin{split}
   M_{11}&=  \frac{8 \sin k-\gamma (1+\delta)}{8 \sin k} \mathbb{I}_2,~~~~M_{12}=\frac{- \gamma(1+\delta)}{8 \sin k} \mathbb{I}_2 \\
   M_{21} &=   \frac{\gamma(1+\delta)}{8 \sin k}\mathbb{I}_2,\qquad\quad M_{22}=\frac{8 \sin k+\gamma(1+\delta)}{8 \sin k}\mathbb{I}_2
    \label{eq:CS_v2_2_sm}
     \end{split}
  \end{equation}
Imposing $ M_{11} (k) = \mathbb{O}_2$ yields the CPA condition 
\begin{equation}
\gamma_* = \frac{8}{1+\delta}\sin k
\quad\Leftrightarrow\quad
k_* = \arcsin \frac{\gamma(1+\delta)}{8}
\label{eq:CS_v2_3_sm}
\end{equation}

\subsection{Cross-Stitch - Nonlinear Regime}\label{sec:cross_stitch_lin}

In the nonlinear regime of the Cross-Stitch lattice we cannot apply the transfer matrix $M$ strategy in Eq.\eqref{eq:cpatm2} to obtain the CPA condition. 
\begin{equation}
\begin{split}
\mu a_n &= - a_{n-1} - a_{n+1} - b_{n-1} - b_{n+1} - h b_{n}+ g |a_n|^2 a_n\\ 
&\quad- \left[ \frac{i\gamma}{2}  a_n +   \frac{i\gamma\delta}{2}b_n\right] \delta_{n,0} \\
\mu b_n &= - b_{n-1} - b_{n+1} - a_{n-1} - a_{n+1} - h a_{n}+ g|b_n|^2 b_n \\
&\quad -  \left[  \frac{i\gamma\delta}{2}a_n + \frac{i\gamma}{2}  b_n \right]\delta_{n,0} 
\end{split}
\label{eq:CS_nonlin_sm}
\end{equation}
Nevertheless, we can show that nonlinearity does not alter the linear CPA condition in the following way.
We consider only the Case 2 - $V_2$ of the dissipation, as this turns into $V_1$ for $\delta = 0$.\\
\\
Let us consider the steady-state solution $\Psi_n = (a_n,b_n)^Te^{-i\mu t}$ for Eq.(\ref{eq:CS_nonlin_sm}).
We then consider the ansatz Eq.(\ref{eq:CS_ansatz_sm}) reduced to $a_L = b_R=\rho \Phi_1$ and $\Phi_1 = \frac{1}{\sqrt{2}}(1,1)^T$ and $b_L=a_R=0$
 \begin{equation}
U_n = \Phi_1
\begin{cases}
                  \rho e^{ikn}  ~~~~~~~~n < 0 \\
                  \rho ~~~~~~~~~~~~~n=0  \\
                 \rho e^{-ikn} ~~~~~~n > 0.
\end{cases}
\label{eq:1d_cs_ansatz2}  
\end{equation}
For small amplitude waves, this fixes $\mu^\prime = -h - 4\cos k + g\rho^2$. 
Eq.(\ref{eq:CS_nonlin_sm}) at $n=0$ then reads
  \begin{equation}
\begin{split}
&\quad \mu^\prime \rho = - 2\rho ( e^{- ik} + e^{-ik}  ) -h \rho + g \rho^3 - i\frac{\gamma}{2} (1 + \delta)\rho\\
\quad & \Leftrightarrow \quad
\mu^\prime  = -4   e^{- ik} -h  + g \rho^2 - i\frac{\gamma}{2}(1 + \delta)
 \\
 \quad & \Leftrightarrow \quad
-h - 4\cos k + g\rho^2 = -4   e^{- ik} -h  + g \rho^2 - i\frac{\gamma}{2}(1 + \delta) \\
  \quad & \Leftrightarrow \quad
 -4 \cos k = -4  e^{- ik} - i\frac{\gamma}{2}(1 + \delta) \\
  \quad & \Leftrightarrow \quad
 \cos k -  e^{- ik} =  i\frac{\gamma}{8}(1 + \delta)\\
  \quad & \Leftrightarrow \quad
\frac{e^{ ik} - e^{- ik}}{2}  =  i\frac{\gamma}{8}(1 + \delta)
  \quad  \Leftrightarrow \quad
\sin k =  \frac{\gamma(1 + \delta)}{8} 
 \end{split}
\label{eq:1d_cs_n=0}
\end{equation}
yielding the same CPA condition Eq.(\ref{eq:CS_v2_3_sm}) valid in the linear regime $g=0$.

\section{numerical simulation} \label{sec:app2}

\subsection{Gaussian beams}

\begin{figure}[H]  
 \centering
\includegraphics[width=0.95\columnwidth]{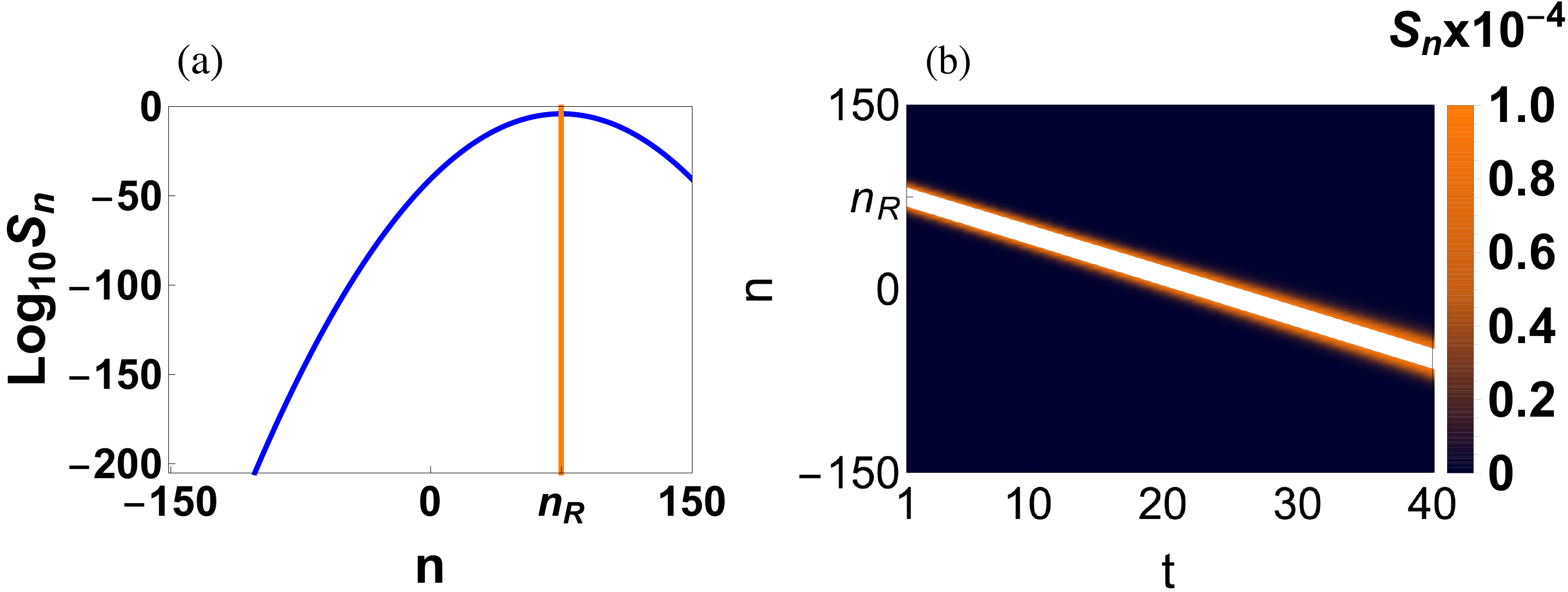}
 \caption{
(a) Spatial profile of the initial excitation in Eq.(\ref{eq:CS_IC_sm}) for $P_0=0.01$ and $\alpha=0.0075$, with $n_R = 75$. 
(b)  Time evolution of $S_n$ for $k=1$. 
Here $g=1$, $h=1$.
}
  \label{fig:app1}
\end{figure}

Let us consider the Cross-Stitch lattice in Eq.(\ref{eq:CS_nonlin_sm}) written in components $\Psi_n = (a_n,b_n)^T$ with no dissipation $\gamma=0$. 
Since the mode $\Phi_1 = \frac{1}{\sqrt{2}}(1,1)^T$  associated to the dispersive band of the model $\mu_1(k) = -h - 4\cos k$ is symmetric upon the two components, 
we initialize a small amplitude propagating wave with Gaussian profile located at the unit-cell $n_R\gg 0$ - as shown in Fig.\ref{fig:app1}(a) - by defining a symmetric excitation $a_n = b_n$, with 
\begin{equation}
\begin{split}
a_n &= P_0 e^{-\alpha (n - n_R)^2}e^{-ik(n - n_R)}
\end{split}
\label{eq:CS_IC_sm}
\end{equation}
Here $k$ is the quasi-momentum, $\alpha$ is called the inverse width of the wave and $P_0$ is the amplitude.
In Fig.\ref{fig:app1}(b) we show the time-evolution of this Gaussian beam. 
Let us observe that this wave packet is not mono-chromatic. However, the choice of $\alpha$ implies a spatial width of the beam of approximatively fifteen unit-cells, with a correspondent frequency range of the packet in the reciprocal $k$-space lower than $10^{-1}$. 
This choice therefore keeps the frequency width of the wave packet close to the desired one. 
%, and consequently it allows to employ Gaussian beams to successfully fine-tune CPA.

\begin{figure}[h]  
 \centering
\includegraphics[width=0.95\columnwidth]{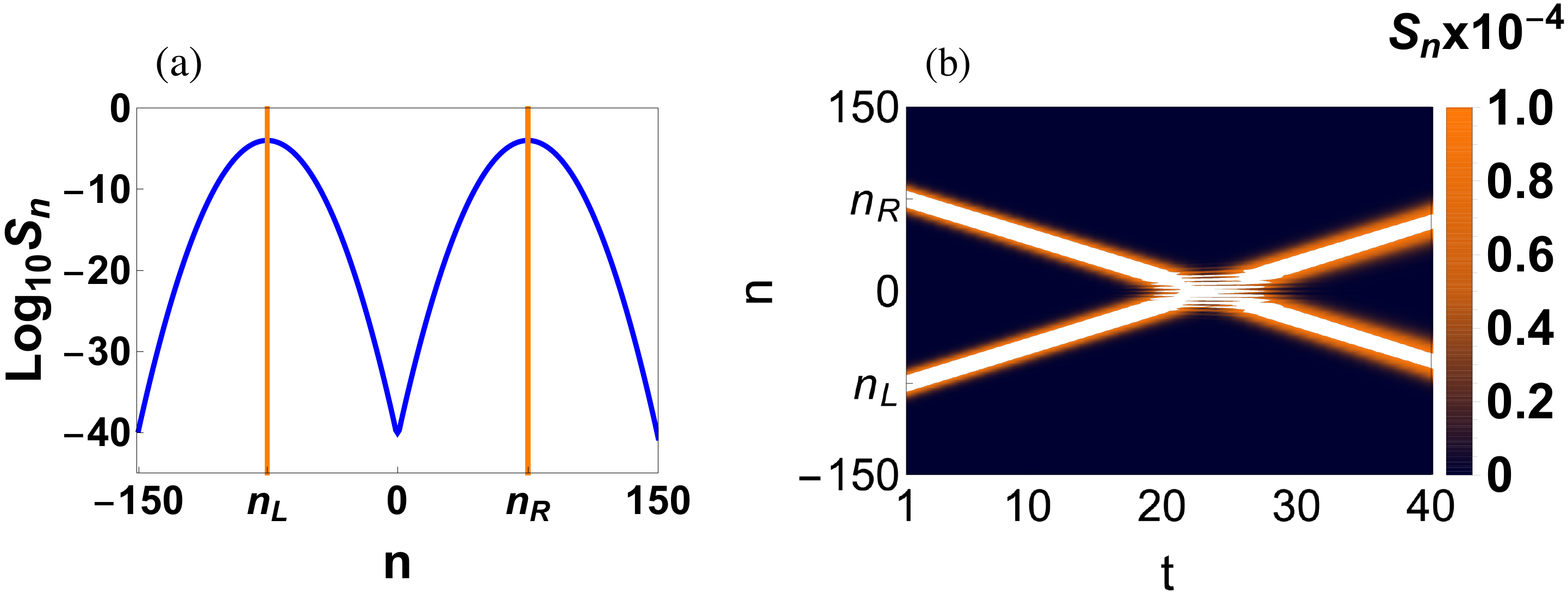}
 \caption{
(a) Spatial profile of the initial excitation in Eq.(\ref{eq:CS_IC2_sm}) for $P=0.01$ and $\alpha=0.0075$, with $n_R = 75$ and $n_L =-75$ . 
(b)  Time evolution of $S_n$ for $k=1$. 
Here $g=1$, $h=1$.
}
  \label{fig:app2}
\end{figure}

In order to achieve CPA, two counter-propagating beams 
which meet at unit-cell $n=0$ (where the dissipation will be located) are required.
We achieve this by considering two equal beams in Eq.(\ref{eq:CS_IC_sm}) which departs from opposite sides $n_R\gg 0$ and $n_L\ll 0$ with $n_L = -n_R$ and opposite quasi-momentum $k$ - as shown in Fig.\ref{fig:app2}(a) - by defining a symmetric excitation $a_n = b_n$ with

\begin{equation}
\begin{split}
a_n &= P_0 e^{-\alpha (n - n_R)^2}e^{-ik(n - n_R)}
+ P_0 e^{-\alpha (n - n_L)^2}e^{ik(n - n_L)} \\
\end{split}
\label{eq:CS_IC2_sm}
\end{equation}
%Note that since left and right beams have  to ensure counter-propagation. 
This is visualized in Fig.\ref{fig:app2}(b), where we see that the two beams collide at the unit-cell $n=0$.

The very same strategy has been applied to visualize CPA in the nonlinear diamond chain, whose components are $\Psi_n = (a_n,b_n,c_n)^T$.
Since the eigenmodes $\phi_{1,2} = (1,1,\pm \sqrt{2} e^{-k/2})^T / 2$
of the two dispersive bands $\mu_{1,2}(k) = (- h \pm \sqrt{h^2+16+16 \cos k})/2$ is symmetric on the outer components $a_n,b_n$ and has $\pm \sqrt{2} e^{-k/2}$ as a multiplicative factor on the central component $c_n$, 
we define the Gaussian beams 
\begin{equation}
\begin{split}
a_n &= P_0 e^{-\alpha (n - n_R)^2}e^{-ik(n - n_R)}
+ P_0 e^{-\alpha (n - n_L)^2}e^{ik(n - n_L)}
\end{split}
\label{eq:DC_IC_sm}
\end{equation}
with $b_n = a_n$ and  $c_n = \pm \sqrt{2} e^{-k/2} a_n$.

\subsection{Computations}

All the numerical simulations of the time-evolution herewith shown have been performed using the commercial software Mathematica and  employing the $4^{th}$-order explicit Runge-Kutta scheme.

\bibliographystyle{apsrev4}
\let\itshape\upshape
\normalem
\bibliography{reference1}

\providecommand{\noopsort}[1]{}\providecommand{\singleletter}[1]{#1}%
\begin{thebibliography}{43}%
\makeatletter
\providecommand \@ifxundefined [1]{%
 \@ifx{#1\undefined}
}%
\providecommand \@ifnum [1]{%
 \ifnum #1\expandafter \@firstoftwo
 \else \expandafter \@secondoftwo
 \fi
}%
\providecommand \@ifx [1]{%
 \ifx #1\expandafter \@firstoftwo
 \else \expandafter \@secondoftwo
 \fi
}%
\providecommand \natexlab [1]{#1}%
\providecommand \enquote  [1]{``#1''}%
\providecommand \bibnamefont  [1]{#1}%
\providecommand \bibfnamefont [1]{#1}%
\providecommand \citenamefont [1]{#1}%
\providecommand \href@noop [0]{\@secondoftwo}%
\providecommand \href [0]{\begingroup \@sanitize@url \@href}%
\providecommand \@href[1]{\@@startlink{#1}\@@href}%
\providecommand \@@href[1]{\endgroup#1\@@endlink}%
\providecommand \@sanitize@url [0]{\catcode `\\12\catcode `\$12\catcode
  `\&12\catcode `\#12\catcode `\^12\catcode `\_12\catcode `\%12\relax}%
\providecommand \@@startlink[1]{}%
\providecommand \@@endlink[0]{}%
\providecommand \url  [0]{\begingroup\@sanitize@url \@url }%
\providecommand \@url [1]{\endgroup\@href {#1}{\urlprefix }}%
\providecommand \urlprefix  [0]{URL }%
\providecommand \Eprint [0]{\href }%
\providecommand \doibase [0]{http://dx.doi.org/}%
\providecommand \selectlanguage [0]{\@gobble}%
\providecommand \bibinfo  [0]{\@secondoftwo}%
\providecommand \bibfield  [0]{\@secondoftwo}%
\providecommand \translation [1]{[#1]}%
\providecommand \BibitemOpen [0]{}%
\providecommand \bibitemStop [0]{}%
\providecommand \bibitemNoStop [0]{.\EOS\space}%
\providecommand \EOS [0]{\spacefactor3000\relax}%
\providecommand \BibitemShut  [1]{\csname bibitem#1\endcsname}%
\let\auto@bib@innerbib\@empty
%</preamble>
\bibitem [{\citenamefont {Mostafazadeh}(2009)}]{Mostafazadeh:2009}%
  \BibitemOpen
  \bibfield  {author} {\bibinfo {author} {\bibfnamefont {A.}~\bibnamefont
  {Mostafazadeh}},\ }\bibfield  {title} {\enquote {\bibinfo {title} {Spectral
  singularities of complex scattering potentials and infinite reflection and
  transmission coefficients at real energies},}\ }\href {\doibase
  10.1103/PhysRevLett.102.220402} {\bibfield  {journal} {\bibinfo  {journal}
  {Phys. Rev. Lett.}\ }\textbf {\bibinfo {volume} {102}},\ \bibinfo {pages}
  {220402} (\bibinfo {year} {2009})}\BibitemShut {NoStop}%
\bibitem [{\citenamefont {Baranov}\ \emph {et~al.}(2017)\citenamefont
  {Baranov}, \citenamefont {Krasnok}, \citenamefont {Shegai}, \citenamefont
  {Al{\`u}},\ and\ \citenamefont {Chong}}]{baranov2017coherent}%
  \BibitemOpen
  \bibfield  {author} {\bibinfo {author} {\bibfnamefont {D.~G.}\ \bibnamefont
  {Baranov}}, \bibinfo {author} {\bibfnamefont {A.}~\bibnamefont {Krasnok}},
  \bibinfo {author} {\bibfnamefont {T.}~\bibnamefont {Shegai}}, \bibinfo
  {author} {\bibfnamefont {A.}~\bibnamefont {Al{\`u}}}, \ and\ \bibinfo
  {author} {\bibfnamefont {Y.}~\bibnamefont {Chong}},\ }\bibfield  {title}
  {\enquote {\bibinfo {title} {Coherent perfect absorbers: linear control of
  light with light},}\ }\href {https://doi.org/10.1038/natrevmats.2017.64}
  {\bibfield  {journal} {\bibinfo  {journal} {Nature Reviews Materials}\
  }\textbf {\bibinfo {volume} {2}},\ \bibinfo {pages} {17064} (\bibinfo {year}
  {2017})}\BibitemShut {NoStop}%
\bibitem [{\citenamefont {Chong}\ \emph {et~al.}(2010)\citenamefont {Chong},
  \citenamefont {Ge}, \citenamefont {Cao},\ and\ \citenamefont
  {Stone}}]{Chong2010coherent}%
  \BibitemOpen
  \bibfield  {author} {\bibinfo {author} {\bibfnamefont {Y.~D.}\ \bibnamefont
  {Chong}}, \bibinfo {author} {\bibfnamefont {L.}~\bibnamefont {Ge}}, \bibinfo
  {author} {\bibfnamefont {H.}~\bibnamefont {Cao}}, \ and\ \bibinfo {author}
  {\bibfnamefont {A.~D.}\ \bibnamefont {Stone}},\ }\bibfield  {title} {\enquote
  {\bibinfo {title} {Coherent perfect absorbers: Time-reversed lasers},}\
  }\href {\doibase 10.1103/PhysRevLett.105.053901} {\bibfield  {journal}
  {\bibinfo  {journal} {Phys. Rev. Lett.}\ }\textbf {\bibinfo {volume} {105}},\
  \bibinfo {pages} {053901} (\bibinfo {year} {2010})}\BibitemShut {NoStop}%
\bibitem [{\citenamefont {Longhi}(2010)}]{Longhi:2010}%
  \BibitemOpen
  \bibfield  {author} {\bibinfo {author} {\bibfnamefont {S.}~\bibnamefont
  {Longhi}},\ }\bibfield  {title} {\enquote {\bibinfo {title}
  {$\mathcal{PT}$-symmetric laser absorber},}\ }\href {\doibase
  10.1103/PhysRevA.82.031801} {\bibfield  {journal} {\bibinfo  {journal} {Phys.
  Rev. A}\ }\textbf {\bibinfo {volume} {82}},\ \bibinfo {pages} {031801}
  (\bibinfo {year} {2010})}\BibitemShut {NoStop}%
\bibitem [{\citenamefont {Chong}\ \emph {et~al.}(2011)\citenamefont {Chong},
  \citenamefont {Ge},\ and\ \citenamefont {Stone}}]{Chong:2011PT}%
  \BibitemOpen
  \bibfield  {author} {\bibinfo {author} {\bibfnamefont {Y.~D.}\ \bibnamefont
  {Chong}}, \bibinfo {author} {\bibfnamefont {L.}~\bibnamefont {Ge}}, \ and\
  \bibinfo {author} {\bibfnamefont {A.~D.}\ \bibnamefont {Stone}},\ }\bibfield
  {title} {\enquote {\bibinfo {title} {$\mathcal{PT}$-symmetry breaking and
  laser-absorber modes in optical scattering systems},}\ }\href {\doibase
  10.1103/PhysRevLett.106.093902} {\bibfield  {journal} {\bibinfo  {journal}
  {Phys. Rev. Lett.}\ }\textbf {\bibinfo {volume} {106}},\ \bibinfo {pages}
  {093902} (\bibinfo {year} {2011})}\BibitemShut {NoStop}%
\bibitem [{\citenamefont {Zezyulin}\ \emph {et~al.}(2018)\citenamefont
  {Zezyulin}, \citenamefont {Ott},\ and\ \citenamefont
  {Konotop}}]{Zezyulin1018array}%
  \BibitemOpen
  \bibfield  {author} {\bibinfo {author} {\bibfnamefont {D.~A.}\ \bibnamefont
  {Zezyulin}}, \bibinfo {author} {\bibfnamefont {H.}~\bibnamefont {Ott}}, \
  and\ \bibinfo {author} {\bibfnamefont {V.~V.}\ \bibnamefont {Konotop}},\
  }\bibfield  {title} {\enquote {\bibinfo {title} {Coherent perfect absorber
  and laser for nonlinear waves in optical waveguide arrays},}\ }\href
  {\doibase 10.1364/OL.43.005901} {\bibfield  {journal} {\bibinfo  {journal}
  {Opt. Lett.}\ }\textbf {\bibinfo {volume} {43}},\ \bibinfo {pages} {5901}
  (\bibinfo {year} {2018})}\BibitemShut {NoStop}%
\bibitem [{\citenamefont {M{\"u}llers}\ \emph {et~al.}(2018)\citenamefont
  {M{\"u}llers}, \citenamefont {Santra}, \citenamefont {Baals}, \citenamefont
  {Jiang}, \citenamefont {Benary}, \citenamefont {Labouvie}, \citenamefont
  {Zezyulin}, \citenamefont {Konotop},\ and\ \citenamefont
  {Ott}}]{Mullerseaat6539}%
  \BibitemOpen
  \bibfield  {author} {\bibinfo {author} {\bibfnamefont {A.}~\bibnamefont
  {M{\"u}llers}}, \bibinfo {author} {\bibfnamefont {B.}~\bibnamefont {Santra}},
  \bibinfo {author} {\bibfnamefont {C.}~\bibnamefont {Baals}}, \bibinfo
  {author} {\bibfnamefont {J.}~\bibnamefont {Jiang}}, \bibinfo {author}
  {\bibfnamefont {J.}~\bibnamefont {Benary}}, \bibinfo {author} {\bibfnamefont
  {R.}~\bibnamefont {Labouvie}}, \bibinfo {author} {\bibfnamefont {D.~A.}\
  \bibnamefont {Zezyulin}}, \bibinfo {author} {\bibfnamefont {V.~V.}\
  \bibnamefont {Konotop}}, \ and\ \bibinfo {author} {\bibfnamefont
  {H.}~\bibnamefont {Ott}},\ }\bibfield  {title} {\enquote {\bibinfo {title}
  {Coherent perfect absorption of nonlinear matter waves},}\ }\href {\doibase
  10.1126/sciadv.aat6539} {\bibfield  {journal} {\bibinfo  {journal} {Science
  Advances}\ }\textbf {\bibinfo {volume} {4}} (\bibinfo {year} {2018}),\
  10.1126/sciadv.aat6539}\BibitemShut {NoStop}%
\bibitem [{\citenamefont {Jeffers}(2019)}]{jeffers2019nonlocal}%
  \BibitemOpen
  \bibfield  {author} {\bibinfo {author} {\bibfnamefont {J.}~\bibnamefont
  {Jeffers}},\ }\bibfield  {title} {\enquote {\bibinfo {title} {Nonlocal
  coherent perfect absorption},}\ }\href {\doibase
  10.1103/PhysRevLett.123.143602} {\bibfield  {journal} {\bibinfo  {journal}
  {Phys. Rev. Lett.}\ }\textbf {\bibinfo {volume} {123}},\ \bibinfo {pages}
  {143602} (\bibinfo {year} {2019})}\BibitemShut {NoStop}%
\bibitem [{\citenamefont {Leykam}\ \emph {et~al.}(2018)\citenamefont {Leykam},
  \citenamefont {Andreanov},\ and\ \citenamefont
  {Flach}}]{Leykam2018artificial}%
  \BibitemOpen
  \bibfield  {author} {\bibinfo {author} {\bibfnamefont {D.}~\bibnamefont
  {Leykam}}, \bibinfo {author} {\bibfnamefont {A.}~\bibnamefont {Andreanov}}, \
  and\ \bibinfo {author} {\bibfnamefont {S.}~\bibnamefont {Flach}},\ }\bibfield
   {title} {\enquote {\bibinfo {title} {Artificial flat band systems: from
  lattice models to experiments},}\ }\href {\doibase
  10.1080/23746149.2018.1473052} {\bibfield  {journal} {\bibinfo  {journal}
  {Advances in Physics: X}\ }\textbf {\bibinfo {volume} {3}},\ \bibinfo {pages}
  {1473052} (\bibinfo {year} {2018})}\BibitemShut {NoStop}%
\bibitem [{\citenamefont {Leykam}\ and\ \citenamefont
  {Flach}(2018)}]{Leykam2018perspective}%
  \BibitemOpen
  \bibfield  {author} {\bibinfo {author} {\bibfnamefont {D.}~\bibnamefont
  {Leykam}}\ and\ \bibinfo {author} {\bibfnamefont {S.}~\bibnamefont {Flach}},\
  }\bibfield  {title} {\enquote {\bibinfo {title} {Perspective: Photonic
  flatbands},}\ }\href {\doibase 10.1063/1.5034365} {\bibfield  {journal}
  {\bibinfo  {journal} {APL Photonics}\ }\textbf {\bibinfo {volume} {3}},\
  \bibinfo {pages} {070901} (\bibinfo {year} {2018})}\BibitemShut {NoStop}%
\bibitem [{\citenamefont {Flach}\ \emph {et~al.}(2014)\citenamefont {Flach},
  \citenamefont {Leykam}, \citenamefont {Bodyfelt}, \citenamefont {Matthies},\
  and\ \citenamefont {Desyatnikov}}]{Flach2014detangling}%
  \BibitemOpen
  \bibfield  {author} {\bibinfo {author} {\bibfnamefont {S.}~\bibnamefont
  {Flach}}, \bibinfo {author} {\bibfnamefont {D.}~\bibnamefont {Leykam}},
  \bibinfo {author} {\bibfnamefont {J.~D.}\ \bibnamefont {Bodyfelt}}, \bibinfo
  {author} {\bibfnamefont {P.}~\bibnamefont {Matthies}}, \ and\ \bibinfo
  {author} {\bibfnamefont {A.~S.}\ \bibnamefont {Desyatnikov}},\ }\bibfield
  {title} {\enquote {\bibinfo {title} {Detangling flat bands into fano
  lattices},}\ }\href {\doibase 10.1209/0295-5075/105/30001} {\bibfield
  {journal} {\bibinfo  {journal} {{EPL} (Europhysics Letters)}\ }\textbf
  {\bibinfo {volume} {105}},\ \bibinfo {pages} {30001} (\bibinfo {year}
  {2014})}\BibitemShut {NoStop}%
\bibitem [{\citenamefont {Maimaiti}\ \emph {et~al.}(2017)\citenamefont
  {Maimaiti}, \citenamefont {Andreanov}, \citenamefont {Park}, \citenamefont
  {Gendelman},\ and\ \citenamefont {Flach}}]{Maimaiti2017compact}%
  \BibitemOpen
  \bibfield  {author} {\bibinfo {author} {\bibfnamefont {W.}~\bibnamefont
  {Maimaiti}}, \bibinfo {author} {\bibfnamefont {A.}~\bibnamefont {Andreanov}},
  \bibinfo {author} {\bibfnamefont {H.~C.}\ \bibnamefont {Park}}, \bibinfo
  {author} {\bibfnamefont {O.}~\bibnamefont {Gendelman}}, \ and\ \bibinfo
  {author} {\bibfnamefont {S.}~\bibnamefont {Flach}},\ }\bibfield  {title}
  {\enquote {\bibinfo {title} {Compact localized states and flat-band
  generators in one dimension},}\ }\href {\doibase 10.1103/PhysRevB.95.115135}
  {\bibfield  {journal} {\bibinfo  {journal} {Phys. Rev. B}\ }\textbf {\bibinfo
  {volume} {95}},\ \bibinfo {pages} {115135} (\bibinfo {year}
  {2017})}\BibitemShut {NoStop}%
\bibitem [{\citenamefont {R\"ontgen}\ \emph {et~al.}(2018)\citenamefont
  {R\"ontgen}, \citenamefont {Morfonios},\ and\ \citenamefont
  {Schmelcher}}]{Rontgen2018compact}%
  \BibitemOpen
  \bibfield  {author} {\bibinfo {author} {\bibfnamefont {M.}~\bibnamefont
  {R\"ontgen}}, \bibinfo {author} {\bibfnamefont {C.~V.}\ \bibnamefont
  {Morfonios}}, \ and\ \bibinfo {author} {\bibfnamefont {P.}~\bibnamefont
  {Schmelcher}},\ }\bibfield  {title} {\enquote {\bibinfo {title} {Compact
  localized states and flat bands from local symmetry partitioning},}\ }\href
  {\doibase 10.1103/PhysRevB.97.035161} {\bibfield  {journal} {\bibinfo
  {journal} {Phys. Rev. B}\ }\textbf {\bibinfo {volume} {97}},\ \bibinfo
  {pages} {035161} (\bibinfo {year} {2018})}\BibitemShut {NoStop}%
\bibitem [{\citenamefont {Maimaiti}\ \emph {et~al.}(2019)\citenamefont
  {Maimaiti}, \citenamefont {Flach},\ and\ \citenamefont
  {Andreanov}}]{Maimaiti2019universal}%
  \BibitemOpen
  \bibfield  {author} {\bibinfo {author} {\bibfnamefont {W.}~\bibnamefont
  {Maimaiti}}, \bibinfo {author} {\bibfnamefont {S.}~\bibnamefont {Flach}}, \
  and\ \bibinfo {author} {\bibfnamefont {A.}~\bibnamefont {Andreanov}},\
  }\bibfield  {title} {\enquote {\bibinfo {title} {Universal $d=1$ flat band
  generator from compact localized states},}\ }\href {\doibase
  10.1103/PhysRevB.99.125129} {\bibfield  {journal} {\bibinfo  {journal} {Phys.
  Rev. B}\ }\textbf {\bibinfo {volume} {99}},\ \bibinfo {pages} {125129}
  (\bibinfo {year} {2019})}\BibitemShut {NoStop}%
\bibitem [{\citenamefont {Leykam}\ \emph {et~al.}(2013)\citenamefont {Leykam},
  \citenamefont {Flach}, \citenamefont {Bahat-Treidel},\ and\ \citenamefont
  {Desyatnikov}}]{Leykam2013flatband}%
  \BibitemOpen
  \bibfield  {author} {\bibinfo {author} {\bibfnamefont {D.}~\bibnamefont
  {Leykam}}, \bibinfo {author} {\bibfnamefont {S.}~\bibnamefont {Flach}},
  \bibinfo {author} {\bibfnamefont {O.}~\bibnamefont {Bahat-Treidel}}, \ and\
  \bibinfo {author} {\bibfnamefont {A.~S.}\ \bibnamefont {Desyatnikov}},\
  }\bibfield  {title} {\enquote {\bibinfo {title} {Flat band states: Disorder
  and nonlinearity},}\ }\href {\doibase 10.1103/PhysRevB.88.224203} {\bibfield
  {journal} {\bibinfo  {journal} {Phys. Rev. B}\ }\textbf {\bibinfo {volume}
  {88}},\ \bibinfo {pages} {224203} (\bibinfo {year} {2013})}\BibitemShut
  {NoStop}%
\bibitem [{\citenamefont {Bodyfelt}\ \emph {et~al.}(2014)\citenamefont
  {Bodyfelt}, \citenamefont {Leykam}, \citenamefont {Danieli}, \citenamefont
  {Yu},\ and\ \citenamefont {Flach}}]{Bodyfelt2014flatbands}%
  \BibitemOpen
  \bibfield  {author} {\bibinfo {author} {\bibfnamefont {J.~D.}\ \bibnamefont
  {Bodyfelt}}, \bibinfo {author} {\bibfnamefont {D.}~\bibnamefont {Leykam}},
  \bibinfo {author} {\bibfnamefont {C.}~\bibnamefont {Danieli}}, \bibinfo
  {author} {\bibfnamefont {X.}~\bibnamefont {Yu}}, \ and\ \bibinfo {author}
  {\bibfnamefont {S.}~\bibnamefont {Flach}},\ }\bibfield  {title} {\enquote
  {\bibinfo {title} {Flatbands under correlated perturbations},}\ }\href
  {\doibase 10.1103/PhysRevLett.113.236403} {\bibfield  {journal} {\bibinfo
  {journal} {Phys. Rev. Lett.}\ }\textbf {\bibinfo {volume} {113}},\ \bibinfo
  {pages} {236403} (\bibinfo {year} {2014})}\BibitemShut {NoStop}%
\bibitem [{\citenamefont {Danieli}\ \emph {et~al.}(2015)\citenamefont
  {Danieli}, \citenamefont {Bodyfelt},\ and\ \citenamefont
  {Flach}}]{Danieli2015flatbands}%
  \BibitemOpen
  \bibfield  {author} {\bibinfo {author} {\bibfnamefont {C.}~\bibnamefont
  {Danieli}}, \bibinfo {author} {\bibfnamefont {J.~D.}\ \bibnamefont
  {Bodyfelt}}, \ and\ \bibinfo {author} {\bibfnamefont {S.}~\bibnamefont
  {Flach}},\ }\bibfield  {title} {\enquote {\bibinfo {title} {Flat-band
  engineering of mobility edges},}\ }\href {\doibase
  10.1103/PhysRevB.91.235134} {\bibfield  {journal} {\bibinfo  {journal} {Phys.
  Rev. B}\ }\textbf {\bibinfo {volume} {91}},\ \bibinfo {pages} {235134}
  (\bibinfo {year} {2015})}\BibitemShut {NoStop}%
\bibitem [{\citenamefont {Leykam}\ \emph
  {et~al.}(2017{\natexlab{a}})\citenamefont {Leykam}, \citenamefont {Bodyfelt},
  \citenamefont {Desyatnikov},\ and\ \citenamefont
  {Flach}}]{Leykam2017localization}%
  \BibitemOpen
  \bibfield  {author} {\bibinfo {author} {\bibfnamefont {D.}~\bibnamefont
  {Leykam}}, \bibinfo {author} {\bibfnamefont {J.~D.}\ \bibnamefont
  {Bodyfelt}}, \bibinfo {author} {\bibfnamefont {A.~S.}\ \bibnamefont
  {Desyatnikov}}, \ and\ \bibinfo {author} {\bibfnamefont {S.}~\bibnamefont
  {Flach}},\ }\bibfield  {title} {\enquote {\bibinfo {title} {Localization of
  weakly disordered flat band states},}\ }\href {\doibase
  10.1140/epjb/e2016-70551-2} {\bibfield  {journal} {\bibinfo  {journal} {The
  European Physical Journal B}\ }\textbf {\bibinfo {volume} {90}},\ \bibinfo
  {pages} {1} (\bibinfo {year} {2017}{\natexlab{a}})}\BibitemShut {NoStop}%
\bibitem [{\citenamefont {Johansson}\ \emph {et~al.}(2015)\citenamefont
  {Johansson}, \citenamefont {Naether},\ and\ \citenamefont
  {Vicencio}}]{Johansson2015compactification}%
  \BibitemOpen
  \bibfield  {author} {\bibinfo {author} {\bibfnamefont {M.}~\bibnamefont
  {Johansson}}, \bibinfo {author} {\bibfnamefont {U.}~\bibnamefont {Naether}},
  \ and\ \bibinfo {author} {\bibfnamefont {R.~A.}\ \bibnamefont {Vicencio}},\
  }\bibfield  {title} {\enquote {\bibinfo {title} {Compactification tuning for
  nonlinear localized modes in sawtooth lattices},}\ }\href {\doibase
  10.1103/PhysRevE.92.032912} {\bibfield  {journal} {\bibinfo  {journal} {Phys.
  Rev. E}\ }\textbf {\bibinfo {volume} {92}},\ \bibinfo {pages} {032912}
  (\bibinfo {year} {2015})}\BibitemShut {NoStop}%
\bibitem [{\citenamefont {Ramachandran}\ \emph {et~al.}(2018)\citenamefont
  {Ramachandran}, \citenamefont {Danieli},\ and\ \citenamefont
  {Flach}}]{Ramachandran2018fano}%
  \BibitemOpen
  \bibfield  {author} {\bibinfo {author} {\bibfnamefont {A.}~\bibnamefont
  {Ramachandran}}, \bibinfo {author} {\bibfnamefont {C.}~\bibnamefont
  {Danieli}}, \ and\ \bibinfo {author} {\bibfnamefont {S.}~\bibnamefont
  {Flach}},\ }\enquote {\bibinfo {title} {Fano resonances in flat band
  networks},}\ in\ \href {\doibase 10.1007/978-3-319-99731-5_13} {\emph
  {\bibinfo {booktitle} {Fano Resonances in Optics and Microwaves: Physics and
  Applications}}},\ \bibinfo {editor} {edited by\ \bibinfo {editor}
  {\bibfnamefont {E.}~\bibnamefont {Kamenetskii}}, \bibinfo {editor}
  {\bibfnamefont {A.}~\bibnamefont {Sadreev}}, \ and\ \bibinfo {editor}
  {\bibfnamefont {A.}~\bibnamefont {Miroshnichenko}}}\ (\bibinfo  {publisher}
  {Springer International Publishing},\ \bibinfo {address} {Cham},\ \bibinfo
  {year} {2018})\ pp.\ \bibinfo {pages} {311--329}\BibitemShut {NoStop}%
\bibitem [{\citenamefont {Danieli}\ \emph {et~al.}(2018)\citenamefont
  {Danieli}, \citenamefont {Maluckov},\ and\ \citenamefont
  {Flach}}]{Danieli2018compact}%
  \BibitemOpen
  \bibfield  {author} {\bibinfo {author} {\bibfnamefont {C.}~\bibnamefont
  {Danieli}}, \bibinfo {author} {\bibfnamefont {A.}~\bibnamefont {Maluckov}}, \
  and\ \bibinfo {author} {\bibfnamefont {S.}~\bibnamefont {Flach}},\ }\bibfield
   {title} {\enquote {\bibinfo {title} {Compact discrete breathers on flat-band
  networks},}\ }\href {\doibase 10.1063/1.5041434} {\bibfield  {journal}
  {\bibinfo  {journal} {Low Temperature Physics}\ }\textbf {\bibinfo {volume}
  {44}},\ \bibinfo {pages} {678} (\bibinfo {year} {2018})}\BibitemShut
  {NoStop}%
\bibitem [{\citenamefont {Mukherjee}\ \emph {et~al.}(2015)\citenamefont
  {Mukherjee}, \citenamefont {Spracklen}, \citenamefont {Choudhury},
  \citenamefont {Goldman}, \citenamefont {\"Ohberg}, \citenamefont
  {Andersson},\ and\ \citenamefont {Thomson}}]{Mukherjee2015observation}%
  \BibitemOpen
  \bibfield  {author} {\bibinfo {author} {\bibfnamefont {S.}~\bibnamefont
  {Mukherjee}}, \bibinfo {author} {\bibfnamefont {A.}~\bibnamefont
  {Spracklen}}, \bibinfo {author} {\bibfnamefont {D.}~\bibnamefont
  {Choudhury}}, \bibinfo {author} {\bibfnamefont {N.}~\bibnamefont {Goldman}},
  \bibinfo {author} {\bibfnamefont {P.}~\bibnamefont {\"Ohberg}}, \bibinfo
  {author} {\bibfnamefont {E.}~\bibnamefont {Andersson}}, \ and\ \bibinfo
  {author} {\bibfnamefont {R.~R.}\ \bibnamefont {Thomson}},\ }\bibfield
  {title} {\enquote {\bibinfo {title} {Observation of a localized flat-band
  state in a photonic lieb lattice},}\ }\href {\doibase
  10.1103/PhysRevLett.114.245504} {\bibfield  {journal} {\bibinfo  {journal}
  {Phys. Rev. Lett.}\ }\textbf {\bibinfo {volume} {114}},\ \bibinfo {pages}
  {245504} (\bibinfo {year} {2015})}\BibitemShut {NoStop}%
\bibitem [{\citenamefont {Vicencio}\ \emph {et~al.}(2015)\citenamefont
  {Vicencio}, \citenamefont {Cantillano}, \citenamefont {Morales-Inostroza},
  \citenamefont {Real}, \citenamefont {Mej\'{\i}a-Cort\'es}, \citenamefont
  {Weimann}, \citenamefont {Szameit},\ and\ \citenamefont
  {Molina}}]{Vicencio2015observation2}%
  \BibitemOpen
  \bibfield  {author} {\bibinfo {author} {\bibfnamefont {R.~A.}\ \bibnamefont
  {Vicencio}}, \bibinfo {author} {\bibfnamefont {C.}~\bibnamefont
  {Cantillano}}, \bibinfo {author} {\bibfnamefont {L.}~\bibnamefont
  {Morales-Inostroza}}, \bibinfo {author} {\bibfnamefont {B.}~\bibnamefont
  {Real}}, \bibinfo {author} {\bibfnamefont {C.}~\bibnamefont
  {Mej\'{\i}a-Cort\'es}}, \bibinfo {author} {\bibfnamefont {S.}~\bibnamefont
  {Weimann}}, \bibinfo {author} {\bibfnamefont {A.}~\bibnamefont {Szameit}}, \
  and\ \bibinfo {author} {\bibfnamefont {M.~I.}\ \bibnamefont {Molina}},\
  }\bibfield  {title} {\enquote {\bibinfo {title} {Observation of localized
  states in lieb photonic lattices},}\ }\href {\doibase
  10.1103/PhysRevLett.114.245503} {\bibfield  {journal} {\bibinfo  {journal}
  {Phys. Rev. Lett.}\ }\textbf {\bibinfo {volume} {114}},\ \bibinfo {pages}
  {245503} (\bibinfo {year} {2015})}\BibitemShut {NoStop}%
\bibitem [{\citenamefont {Weimann}\ \emph {et~al.}(2016)\citenamefont
  {Weimann}, \citenamefont {Morales-Inostroza}, \citenamefont {Real},
  \citenamefont {Cantillano}, \citenamefont {Szameit},\ and\ \citenamefont
  {Vicencio}}]{Weimann2016transport}%
  \BibitemOpen
  \bibfield  {author} {\bibinfo {author} {\bibfnamefont {S.}~\bibnamefont
  {Weimann}}, \bibinfo {author} {\bibfnamefont {L.}~\bibnamefont
  {Morales-Inostroza}}, \bibinfo {author} {\bibfnamefont {B.}~\bibnamefont
  {Real}}, \bibinfo {author} {\bibfnamefont {C.}~\bibnamefont {Cantillano}},
  \bibinfo {author} {\bibfnamefont {A.}~\bibnamefont {Szameit}}, \ and\
  \bibinfo {author} {\bibfnamefont {R.~A.}\ \bibnamefont {Vicencio}},\
  }\bibfield  {title} {\enquote {\bibinfo {title} {Transport in sawtooth
  photonic lattices},}\ }\href {\doibase 10.1364/OL.41.002414} {\bibfield
  {journal} {\bibinfo  {journal} {Opt. Lett.}\ }\textbf {\bibinfo {volume}
  {41}},\ \bibinfo {pages} {2414} (\bibinfo {year} {2016})}\BibitemShut
  {NoStop}%
\bibitem [{\citenamefont {Masumoto}\ \emph {et~al.}(2012)\citenamefont
  {Masumoto}, \citenamefont {Kim}, \citenamefont {Byrnes}, \citenamefont
  {Kusudo}, \citenamefont {L\"offler}, \citenamefont {H\"ofling}, \citenamefont
  {Forchel},\ and\ \citenamefont {Yamamoto}}]{Masumoto2012exciton}%
  \BibitemOpen
  \bibfield  {author} {\bibinfo {author} {\bibfnamefont {N.}~\bibnamefont
  {Masumoto}}, \bibinfo {author} {\bibfnamefont {N.~Y.}\ \bibnamefont {Kim}},
  \bibinfo {author} {\bibfnamefont {T.}~\bibnamefont {Byrnes}}, \bibinfo
  {author} {\bibfnamefont {K.}~\bibnamefont {Kusudo}}, \bibinfo {author}
  {\bibfnamefont {A.}~\bibnamefont {L\"offler}}, \bibinfo {author}
  {\bibfnamefont {S.}~\bibnamefont {H\"ofling}}, \bibinfo {author}
  {\bibfnamefont {A.}~\bibnamefont {Forchel}}, \ and\ \bibinfo {author}
  {\bibfnamefont {Y.}~\bibnamefont {Yamamoto}},\ }\bibfield  {title} {\enquote
  {\bibinfo {title} {Exciton{\textendash}polariton condensates with flat bands
  in a two-dimensional kagome lattice},}\ }\href {\doibase
  10.1088/1367-2630/14/6/065002} {\bibfield  {journal} {\bibinfo  {journal}
  {New Journal of Physics}\ }\textbf {\bibinfo {volume} {14}},\ \bibinfo
  {pages} {065002} (\bibinfo {year} {2012})}\BibitemShut {NoStop}%
\bibitem [{\citenamefont {Taie}\ \emph {et~al.}(2015)\citenamefont {Taie},
  \citenamefont {Ozawa}, \citenamefont {Ichinose}, \citenamefont {Nishio},
  \citenamefont {Nakajima},\ and\ \citenamefont
  {Takahashi}}]{Taie2015coherent}%
  \BibitemOpen
  \bibfield  {author} {\bibinfo {author} {\bibfnamefont {S.}~\bibnamefont
  {Taie}}, \bibinfo {author} {\bibfnamefont {H.}~\bibnamefont {Ozawa}},
  \bibinfo {author} {\bibfnamefont {T.}~\bibnamefont {Ichinose}}, \bibinfo
  {author} {\bibfnamefont {T.}~\bibnamefont {Nishio}}, \bibinfo {author}
  {\bibfnamefont {S.}~\bibnamefont {Nakajima}}, \ and\ \bibinfo {author}
  {\bibfnamefont {Y.}~\bibnamefont {Takahashi}},\ }\bibfield  {title} {\enquote
  {\bibinfo {title} {Coherent driving and freezing of bosonic matter wave in an
  optical lieb lattice},}\ }\href {\doibase 10.1126/sciadv.1500854} {\bibfield
  {journal} {\bibinfo  {journal} {Science Advances}\ }\textbf {\bibinfo
  {volume} {1}} (\bibinfo {year} {2015}),\ 10.1126/sciadv.1500854}\BibitemShut
  {NoStop}%
\bibitem [{Sup()}]{Supple}%
  \BibitemOpen
  \href@noop {} {\bibinfo  {journal} {See Supplemental Material at [URL will be
  inserted by publisher] for additional information}\ }\BibitemShut {NoStop}%
\bibitem [{\citenamefont {Vicencio}\ \emph {et~al.}(2007)\citenamefont
  {Vicencio}, \citenamefont {Brand},\ and\ \citenamefont
  {Flach}}]{Vicencio2007fano}%
  \BibitemOpen
\bibfield  {journal} {  }\bibfield  {author} {\bibinfo {author} {\bibfnamefont
  {R.~A.}\ \bibnamefont {Vicencio}}, \bibinfo {author} {\bibfnamefont
  {J.}~\bibnamefont {Brand}}, \ and\ \bibinfo {author} {\bibfnamefont
  {S.}~\bibnamefont {Flach}},\ }\bibfield  {title} {\enquote {\bibinfo {title}
  {Fano blockade by a bose-einstein condensate in an optical lattice},}\ }\href
  {\doibase 10.1103/PhysRevLett.98.184102} {\bibfield  {journal} {\bibinfo
  {journal} {Phys. Rev. Lett.}\ }\textbf {\bibinfo {volume} {98}},\ \bibinfo
  {pages} {184102} (\bibinfo {year} {2007})}\BibitemShut {NoStop}%
\bibitem [{\citenamefont {Leykam}\ \emph
  {et~al.}(2017{\natexlab{b}})\citenamefont {Leykam}, \citenamefont {Flach},\
  and\ \citenamefont {Chong}}]{Leykam2017flatbandNH}%
  \BibitemOpen
  \bibfield  {author} {\bibinfo {author} {\bibfnamefont {D.}~\bibnamefont
  {Leykam}}, \bibinfo {author} {\bibfnamefont {S.}~\bibnamefont {Flach}}, \
  and\ \bibinfo {author} {\bibfnamefont {Y.~D.}\ \bibnamefont {Chong}},\
  }\bibfield  {title} {\enquote {\bibinfo {title} {Flat bands in lattices with
  non-hermitian coupling},}\ }\href {\doibase 10.1103/PhysRevB.96.064305}
  {\bibfield  {journal} {\bibinfo  {journal} {Phys. Rev. B}\ }\textbf {\bibinfo
  {volume} {96}},\ \bibinfo {pages} {064305} (\bibinfo {year}
  {2017}{\natexlab{b}})}\BibitemShut {NoStop}%
\bibitem [{\citenamefont {Vidal}\ \emph {et~al.}(2000)\citenamefont {Vidal},
  \citenamefont {Dou\ifmmode~\mbox{\c{c}}\else \c{c}\fi{}ot}, \citenamefont
  {Mosseri},\ and\ \citenamefont {Butaud}}]{Vidal2000interaction}%
  \BibitemOpen
  \bibfield  {author} {\bibinfo {author} {\bibfnamefont {J.}~\bibnamefont
  {Vidal}}, \bibinfo {author} {\bibfnamefont {B.}~\bibnamefont
  {Dou\ifmmode~\mbox{\c{c}}\else \c{c}\fi{}ot}}, \bibinfo {author}
  {\bibfnamefont {R.}~\bibnamefont {Mosseri}}, \ and\ \bibinfo {author}
  {\bibfnamefont {P.}~\bibnamefont {Butaud}},\ }\bibfield  {title} {\enquote
  {\bibinfo {title} {Interaction induced delocalization for two particles in a
  periodic potential},}\ }\href {\doibase 10.1103/PhysRevLett.85.3906}
  {\bibfield  {journal} {\bibinfo  {journal} {Phys. Rev. Lett.}\ }\textbf
  {\bibinfo {volume} {85}},\ \bibinfo {pages} {3906} (\bibinfo {year}
  {2000})}\BibitemShut {NoStop}%
\bibitem [{\citenamefont {Di~Liberto}\ \emph {et~al.}(2019)\citenamefont
  {Di~Liberto}, \citenamefont {Mukherjee},\ and\ \citenamefont
  {Goldman}}]{Diliberto2018nonlinear}%
  \BibitemOpen
  \bibfield  {author} {\bibinfo {author} {\bibfnamefont {M.}~\bibnamefont
  {Di~Liberto}}, \bibinfo {author} {\bibfnamefont {S.}~\bibnamefont
  {Mukherjee}}, \ and\ \bibinfo {author} {\bibfnamefont {N.}~\bibnamefont
  {Goldman}},\ }\bibfield  {title} {\enquote {\bibinfo {title} {Nonlinear
  dynamics of aharonov-bohm cages},}\ }\href {\doibase
  10.1103/PhysRevA.100.043829} {\bibfield  {journal} {\bibinfo  {journal}
  {Phys. Rev. A}\ }\textbf {\bibinfo {volume} {100}},\ \bibinfo {pages}
  {043829} (\bibinfo {year} {2019})}\BibitemShut {NoStop}%
\bibitem [{\citenamefont {Gligori\ifmmode~\acute{c}\else \'{c}\fi{}}\ \emph
  {et~al.}(2019)\citenamefont {Gligori\ifmmode~\acute{c}\else \'{c}\fi{}},
  \citenamefont {Beli\ifmmode~\check{c}\else \v{c}\fi{}ev}, \citenamefont
  {Leykam},\ and\ \citenamefont {Maluckov}}]{Grigoric1029nonlinear}%
  \BibitemOpen
  \bibfield  {author} {\bibinfo {author} {\bibfnamefont {G.}~\bibnamefont
  {Gligori\ifmmode~\acute{c}\else \'{c}\fi{}}}, \bibinfo {author}
  {\bibfnamefont {P.~P.}\ \bibnamefont {Beli\ifmmode~\check{c}\else
  \v{c}\fi{}ev}}, \bibinfo {author} {\bibfnamefont {D.}~\bibnamefont {Leykam}},
  \ and\ \bibinfo {author} {\bibfnamefont {A.}~\bibnamefont {Maluckov}},\
  }\bibfield  {title} {\enquote {\bibinfo {title} {Nonlinear symmetry breaking
  of aharonov-bohm cages},}\ }\href {\doibase 10.1103/PhysRevA.99.013826}
  {\bibfield  {journal} {\bibinfo  {journal} {Phys. Rev. A}\ }\textbf {\bibinfo
  {volume} {99}},\ \bibinfo {pages} {013826} (\bibinfo {year}
  {2019})}\BibitemShut {NoStop}%
\bibitem [{\citenamefont {Khomeriki}\ and\ \citenamefont
  {Flach}(2016)}]{Khomeriki2016landau}%
  \BibitemOpen
  \bibfield  {author} {\bibinfo {author} {\bibfnamefont {R.}~\bibnamefont
  {Khomeriki}}\ and\ \bibinfo {author} {\bibfnamefont {S.}~\bibnamefont
  {Flach}},\ }\bibfield  {title} {\enquote {\bibinfo {title} {Landau-zener
  bloch oscillations with perturbed flat bands},}\ }\href {\doibase
  10.1103/PhysRevLett.116.245301} {\bibfield  {journal} {\bibinfo  {journal}
  {Phys. Rev. Lett.}\ }\textbf {\bibinfo {volume} {116}},\ \bibinfo {pages}
  {245301} (\bibinfo {year} {2016})}\BibitemShut {NoStop}%
\bibitem [{\citenamefont {Mukherjee}\ and\ \citenamefont
  {Thomson}(2015)}]{Mukherjee2015Observation2}%
  \BibitemOpen
  \bibfield  {author} {\bibinfo {author} {\bibfnamefont {S.}~\bibnamefont
  {Mukherjee}}\ and\ \bibinfo {author} {\bibfnamefont {R.~R.}\ \bibnamefont
  {Thomson}},\ }\bibfield  {title} {\enquote {\bibinfo {title} {Observation of
  localized flat-band modes in a quasi-one-dimensional photonic rhombic
  lattice},}\ }\href {\doibase 10.1364/OL.40.005443} {\bibfield  {journal}
  {\bibinfo  {journal} {Opt. Lett.}\ }\textbf {\bibinfo {volume} {40}},\
  \bibinfo {pages} {5443} (\bibinfo {year} {2015})}\BibitemShut {NoStop}%
\bibitem [{\citenamefont {Mukherjee}\ and\ \citenamefont
  {Thomson}(2017)}]{Mukherjee2017Observation}%
  \BibitemOpen
  \bibfield  {author} {\bibinfo {author} {\bibfnamefont {S.}~\bibnamefont
  {Mukherjee}}\ and\ \bibinfo {author} {\bibfnamefont {R.~R.}\ \bibnamefont
  {Thomson}},\ }\bibfield  {title} {\enquote {\bibinfo {title} {Observation of
  robust flat-band localization in driven photonic rhombic lattices},}\ }\href
  {\doibase 10.1364/OL.42.002243} {\bibfield  {journal} {\bibinfo  {journal}
  {Opt. Lett.}\ }\textbf {\bibinfo {volume} {42}},\ \bibinfo {pages} {2243}
  (\bibinfo {year} {2017})}\BibitemShut {NoStop}%
\bibitem [{\citenamefont {Mukherjee}\ \emph {et~al.}(2018)\citenamefont
  {Mukherjee}, \citenamefont {Di~Liberto}, \citenamefont {\"Ohberg},
  \citenamefont {Thomson},\ and\ \citenamefont
  {Goldman}}]{Mukherjee2018experimental}%
  \BibitemOpen
  \bibfield  {author} {\bibinfo {author} {\bibfnamefont {S.}~\bibnamefont
  {Mukherjee}}, \bibinfo {author} {\bibfnamefont {M.}~\bibnamefont
  {Di~Liberto}}, \bibinfo {author} {\bibfnamefont {P.}~\bibnamefont
  {\"Ohberg}}, \bibinfo {author} {\bibfnamefont {R.~R.}\ \bibnamefont
  {Thomson}}, \ and\ \bibinfo {author} {\bibfnamefont {N.}~\bibnamefont
  {Goldman}},\ }\bibfield  {title} {\enquote {\bibinfo {title} {Experimental
  observation of aharonov-bohm cages in photonic lattices},}\ }\href {\doibase
  10.1103/PhysRevLett.121.075502} {\bibfield  {journal} {\bibinfo  {journal}
  {Phys. Rev. Lett.}\ }\textbf {\bibinfo {volume} {121}},\ \bibinfo {pages}
  {075502} (\bibinfo {year} {2018})}\BibitemShut {NoStop}%
\bibitem [{\citenamefont {Ni\ifmmode \mbox{\c{t}}\else
  \c{t}\fi{}\ifmmode~\u{a}\else \u{a}\fi{}}\ \emph {et~al.}(2013)\citenamefont
  {Ni\ifmmode \mbox{\c{t}}\else \c{t}\fi{}\ifmmode~\u{a}\else \u{a}\fi{}},
  \citenamefont {Ostahie},\ and\ \citenamefont {Aldea}}]{Nita2013spectral}%
  \BibitemOpen
  \bibfield  {author} {\bibinfo {author} {\bibfnamefont {M.}~\bibnamefont
  {Ni\ifmmode \mbox{\c{t}}\else \c{t}\fi{}\ifmmode~\u{a}\else \u{a}\fi{}}},
  \bibinfo {author} {\bibfnamefont {B.}~\bibnamefont {Ostahie}}, \ and\
  \bibinfo {author} {\bibfnamefont {A.}~\bibnamefont {Aldea}},\ }\bibfield
  {title} {\enquote {\bibinfo {title} {Spectral and transport properties of the
  two-dimensional lieb lattice},}\ }\href {\doibase 10.1103/PhysRevB.87.125428}
  {\bibfield  {journal} {\bibinfo  {journal} {Phys. Rev. B}\ }\textbf {\bibinfo
  {volume} {87}},\ \bibinfo {pages} {125428} (\bibinfo {year}
  {2013})}\BibitemShut {NoStop}%
\bibitem [{\citenamefont {Eckmann}\ and\ \citenamefont
  {Wayne}(2019)}]{Eckmann2019decay}%
  \BibitemOpen
  \bibfield  {author} {\bibinfo {author} {\bibfnamefont {J.-P.}\ \bibnamefont
  {Eckmann}}\ and\ \bibinfo {author} {\bibfnamefont {C.~E.}\ \bibnamefont
  {Wayne}},\ }\bibfield  {title} {\enquote {\bibinfo {title} {Decay of
  hamiltonian breathers under dissipation},}\ }\href
  {https://arxiv.org/pdf/1907.12632.pdf} {\bibfield  {journal} {\bibinfo
  {journal} {arXiv:1907.12632}\ } (\bibinfo {year} {2019})}\BibitemShut
  {NoStop}%
\bibitem [{\citenamefont {Rodrigues}\ \emph {et~al.}(2012)\citenamefont
  {Rodrigues}, \citenamefont {Li}, \citenamefont {Achilleos}, \citenamefont
  {Kevrekidis}, \citenamefont {Frantzeskakis},\ and\ \citenamefont
  {Bender}}]{rodrigues2012pt}%
  \BibitemOpen
  \bibfield  {author} {\bibinfo {author} {\bibfnamefont {A.}~\bibnamefont
  {Rodrigues}}, \bibinfo {author} {\bibfnamefont {K.}~\bibnamefont {Li}},
  \bibinfo {author} {\bibfnamefont {V.}~\bibnamefont {Achilleos}}, \bibinfo
  {author} {\bibfnamefont {P.}~\bibnamefont {Kevrekidis}}, \bibinfo {author}
  {\bibfnamefont {D.}~\bibnamefont {Frantzeskakis}}, \ and\ \bibinfo {author}
  {\bibfnamefont {C.~M.}\ \bibnamefont {Bender}},\ }\bibfield  {title}
  {\enquote {\bibinfo {title} {$\mathcal{PT}$-symmetric double well potentials
  revisited: bifurcations, stability and dynamics},}\ }\href
  {https://arxiv.org/abs/1207.1066} {\bibfield  {journal} {\bibinfo  {journal}
  {arXiv:1207.1066}\ } (\bibinfo {year} {2012})}\BibitemShut {NoStop}%
\bibitem [{\citenamefont {Hsu}\ \emph {et~al.}(2016)\citenamefont {Hsu},
  \citenamefont {Zhen}, \citenamefont {Stone}, \citenamefont {Joannopoulos},\
  and\ \citenamefont {Soljacic}}]{Hsu2016bound}%
  \BibitemOpen
  \bibfield  {author} {\bibinfo {author} {\bibfnamefont {C.~W.}\ \bibnamefont
  {Hsu}}, \bibinfo {author} {\bibfnamefont {B.}~\bibnamefont {Zhen}}, \bibinfo
  {author} {\bibfnamefont {D.~A.}\ \bibnamefont {Stone}}, \bibinfo {author}
  {\bibfnamefont {J.~D.}\ \bibnamefont {Joannopoulos}}, \ and\ \bibinfo
  {author} {\bibfnamefont {M.}~\bibnamefont {Soljacic}},\ }\bibfield  {title}
  {\enquote {\bibinfo {title} {Bound states in the continuum},}\ }\href
  {\doibase 10.1038/natrevmats.2016.48} {\bibfield  {journal} {\bibinfo
  {journal} {Nature Reviews Materials}\ }\textbf {\bibinfo {volume} {1}},\
  \bibinfo {pages} {16048} (\bibinfo {year} {2016})}\BibitemShut {NoStop}%
\bibitem [{\citenamefont {Fang}\ \emph {et~al.}(2014)\citenamefont {Fang},
  \citenamefont {Lun~Tseng}, \citenamefont {Ou}, \citenamefont {MacDonald},
  \citenamefont {Ping~Tsai},\ and\ \citenamefont {Zheludev}}]{Fang2014fang}%
  \BibitemOpen
  \bibfield  {author} {\bibinfo {author} {\bibfnamefont {X.}~\bibnamefont
  {Fang}}, \bibinfo {author} {\bibfnamefont {M.}~\bibnamefont {Lun~Tseng}},
  \bibinfo {author} {\bibfnamefont {J.-Y.}\ \bibnamefont {Ou}}, \bibinfo
  {author} {\bibfnamefont {K.~F.}\ \bibnamefont {MacDonald}}, \bibinfo {author}
  {\bibfnamefont {D.}~\bibnamefont {Ping~Tsai}}, \ and\ \bibinfo {author}
  {\bibfnamefont {N.~I.}\ \bibnamefont {Zheludev}},\ }\bibfield  {title}
  {\enquote {\bibinfo {title} {Ultrafast all-optical switching via coherent
  modulation of metamaterial absorption},}\ }\href {\doibase 10.1063/1.4870635}
  {\bibfield  {journal} {\bibinfo  {journal} {Applied Physics Letters}\
  }\textbf {\bibinfo {volume} {104}},\ \bibinfo {pages} {141102} (\bibinfo
  {year} {2014})}\BibitemShut {NoStop}%
\bibitem [{\citenamefont {Wan}\ \emph {et~al.}(2011)\citenamefont {Wan},
  \citenamefont {Chong}, \citenamefont {Ge}, \citenamefont {Noh}, \citenamefont
  {Stone},\ and\ \citenamefont {Cao}}]{Wan2011time}%
  \BibitemOpen
  \bibfield  {author} {\bibinfo {author} {\bibfnamefont {W.}~\bibnamefont
  {Wan}}, \bibinfo {author} {\bibfnamefont {Y.}~\bibnamefont {Chong}}, \bibinfo
  {author} {\bibfnamefont {L.}~\bibnamefont {Ge}}, \bibinfo {author}
  {\bibfnamefont {H.}~\bibnamefont {Noh}}, \bibinfo {author} {\bibfnamefont
  {A.~D.}\ \bibnamefont {Stone}}, \ and\ \bibinfo {author} {\bibfnamefont
  {H.}~\bibnamefont {Cao}},\ }\bibfield  {title} {\enquote {\bibinfo {title}
  {Time-reversed lasing and interferometric control of absorption},}\ }\href
  {\doibase 10.1126/science.1200735} {\bibfield  {journal} {\bibinfo  {journal}
  {Science}\ }\textbf {\bibinfo {volume} {331}},\ \bibinfo {pages} {889}
  (\bibinfo {year} {2011})}\BibitemShut {NoStop}%
\bibitem [{\citenamefont {Papaioannou}\ \emph {et~al.}(2016)\citenamefont
  {Papaioannou}, \citenamefont {Plum}, \citenamefont {Valente}, \citenamefont
  {Rogers},\ and\ \citenamefont {Zheludev}}]{Papaioannou2016all}%
  \BibitemOpen
  \bibfield  {author} {\bibinfo {author} {\bibfnamefont {M.}~\bibnamefont
  {Papaioannou}}, \bibinfo {author} {\bibfnamefont {E.}~\bibnamefont {Plum}},
  \bibinfo {author} {\bibfnamefont {J.}~\bibnamefont {Valente}}, \bibinfo
  {author} {\bibfnamefont {E.~T.~F.}\ \bibnamefont {Rogers}}, \ and\ \bibinfo
  {author} {\bibfnamefont {N.~I.}\ \bibnamefont {Zheludev}},\ }\bibfield
  {title} {\enquote {\bibinfo {title} {All-optical multichannel logic based on
  coherent perfect absorption in a plasmonic metamaterial},}\ }\href {\doibase
  10.1063/1.4966269} {\bibfield  {journal} {\bibinfo  {journal} {APL
  Photonics}\ }\textbf {\bibinfo {volume} {1}},\ \bibinfo {pages} {090801}
  (\bibinfo {year} {2016})}\BibitemShut {NoStop}%
\end{thebibliography}%

\end{document}